\documentclass[conference]{IEEEtran}
\usepackage[utf8]{inputenc}
\usepackage{bm}
\usepackage{epsfig,amsfonts,amsbsy,bm,mathrsfs}
\usepackage{amssymb,amsmath,amsthm,latexsym,amscd,amsfonts}
\usepackage{lettrine}
\usepackage{authblk}
\usepackage{graphics}
\usepackage{graphicx}
\usepackage{psfrag,float}
\usepackage{pstricks}
\usepackage{pst-plot}
\usepackage{cite}
\usepackage{balance}
\usepackage{epsfig}
\usepackage{epstopdf}
\usepackage{bbm}
\usepackage{enumitem}
\usepackage{dsfont}
\usepackage{setspace}
\usepackage{float}
\usepackage{relsize}
\usepackage{comment}
\usepackage{subfigure} % For subfigures
\usepackage{algorithm}
\usepackage[noend]{algpseudocode}
\usepackage[nolist]{acronym}
\usepackage{tabularx}
\usepackage{pifont}% http://ctan.org/pkg/pifont
\usepackage{units}

\usepackage{tikz}
\usetikzlibrary{calc}
\usepackage{xcolor}
\usepackage{tabularx}
\usepackage{colortbl}
\usepackage{pgfplots}
\pgfplotsset{compat=newest}
\usetikzlibrary{plotmarks}
\usetikzlibrary{arrows.meta}
\usepgfplotslibrary{patchplots}
\usepackage{grffile}
\newlength\fheight 
\newlength\fwidth 
\usepgfplotslibrary{fillbetween}

\newtheorem{remark}{Remark}

\makeatletter
\newcommand{\gettikzxy}[3]{%
  \tikz@scan@one@point\pgfutil@firstofone#1\relax
  \edef#2{\the\pgf@x}%
  \edef#3{\the\pgf@y}%
}
\definecolor{mygray}{gray}{0.6}

\newcommand{\FF}{ \boldsymbol{F} }

\newcommand{\complexset}[2]{ \mathbb{C}^{#1 \times #2}  }
\newcommand{\realset}[2]{ \mathbb{R}^{#1 \times #2}  }

\begin{acronym}[ACRONYM]
\acro{AI}{artificial intelligence}
\acro{AoA}{angle-of-arrival}
\acro{AoD}{angle-of-departure}
\acro{AWGN}{additive white Gaussian noise}
\acro{BS}{base station}
\acro{BP}{belief propagation}
\acro{CDF}{cumulative density function}
\acro{CFO}{carrier frequency offset}
\acro{CP}{cyclic prefix}
\acro{CRB}{Cram\'er-Rao bound}
\acro{C-V2X}{cellular vehicle-to-anything}
\acro{DA}{data association}
\acro{D-MIMO}{distributed multiple-input multiple-output}
\acro{DL}{downlink}
\acro{DSRC}{dedicated short-range communications}
\acro{EKF}{extended Kalman filter}
\acro{EM}{electromagnetic}
\acro{FIM}{Fisher information matrix}
\acro{GDOP}{geometric dilution of precision}
\acro{GNSS}{global navigation satellite system}
\acro{GPS}{global positioning system}
\acro{IP}{incidence point}
\acro{IQ}{in-phase and quadrature}
\acro{ISAC}{integrated sensing and communication}
\acro{ICI}{inter-carrier interference}
\acro{JCS}{Joint Communication and Sensing}
\acro{JRC}{joint radar and communication}
\acro{JRC2LS}{joint radar communication, computation, localization, and sensing}
\acro{IMU}{inertial measurement unit}
\acro{IOO}{indoor open office}
\acro{IoT}{Internet of Things}
\acro{IRN}{infrastructure reference node}
\acro{KPI}{key performance indicator}
\acro{LoS}{line-of-sight}
\acro{LS}{least-squares}
\acro{MCRB}{misspecified Cram\'er-Rao bound}
\acro{MIMO}{multiple-input multiple-output}
\acro{ML}{maximum likelihood}
\acro{mmWave}{millimeter-wave}
\acro{NLoS}{non-line-of-sight}
\acro{NR}{new radio}
\acro{OFDM}{orthogonal frequency-division multiplexing}
\acro{OTFS}{orthogonal time-frequency-space}
\acro{OEB}{orientation error bound}
\acro{PEB}{position error bound}
\acro{VEB}{velocity error bound}
\acro{PRS}{positioning reference signal}
\acro{QoS}{Quality of Service}
\acro{RAN}{radio access network}
\acro{RAT}{radio access technology}
\acro{RCS}{radar cross section}
\acro{RedCap}{reduced capacity}
\acro{RF}{radio frequency}
\acro{RIS}{reconfigurable intelligent surface}
\acro{RFS}{random finite set}
\acro{RMSE}{root mean squared error}
\acro{RSU}{road-side unit}
\acro{RTK}{real-time kinematic}
\acro{RTT}{round-trip-time}
\acro{SLAM}{simultaneous localization and mapping}
\acro{SLAT}{simultaneous localization and tracking}
\acro{SNR}{signal-to-noise ratio}
\acro{ToA}{time-of-arrival}
\acro{TDoA}{time-difference-of-arrival}
\acro{TR}{time-reversal}
\acro{TXRX}[TX/RX]{transmitter/receiver}
\acro{Tx}{transmitter}
\acro{Rx}{receiver}
\acro{UE}{user equipment}
\acro{UL}{uplink}
\acro{UWB}{ultra wideband}
\acro{V2I}{vehicle-to-infrastructure}
\acro{V2X}{vehicle-to-anything}
\acro{V2V}{vehicle-to-vehicle}
\acro{XL-MIMO}{extra-large MIMO}
\acro{REB}{range error bound}
\end{acronym}
\begin{document}
\bibliographystyle{IEEEtran}
\bstctlcite{IEEEexample:BSTcontrol}
\title{Analysis of V2X Sidelink Positioning in sub-6 GHz}
\author{Yu Ge\IEEEauthorrefmark{1}, Maximilian Stark\IEEEauthorrefmark{2}, Musa Furkan Keskin\IEEEauthorrefmark{1}, Frank Hofmann\IEEEauthorrefmark{2}, Thomas Hansen\IEEEauthorrefmark{2}, Henk Wymeersch\IEEEauthorrefmark{1}\\
\IEEEauthorrefmark{1}Chalmers University of Technology, Sweden
\IEEEauthorrefmark{2}Robert Bosch GmbH, Germany}

\maketitle
\begin{abstract}
Radio positioning is an important part of joint communication and sensing  in beyond 5G communication systems. Existing works mainly focus on the mmWave bands and under-utilize the  sub-6 GHz bands, even though it is  promising for accurate positioning, especially when the multipath is uncomplicated, and meaningful in several important use cases. In this paper, we analyze V2X sidelink positioning and propose a new performance bound that can predict the positioning performance in the presence of severe multipath. Simulation results using ray-tracing data demonstrate the possibility of sidelink positioning, and the efficacy of the new performance bound and its relation with the complexity of the multipath.   
\end{abstract}
\section{Introduction}
Sensing and positioning have gained significant attention in the evolution of 5G mobile radio systems \cite{wild2021joint}. Much of this attention has been devoted to the mmWave bands (both lower mmWave around 28 GHz and upper mmWave around 140 GHz), due to larger available bandwidth and commensurate distance resolution \cite{KanRap:21,hexax_d31}. Nevertheless, sub-6 GHz bands hold significant promise for accurate positioning as well, especially in scenarios without too complicated multipath.  Due to the presence of the ITS 5850-5925 MHz band and the neighboring unlicensed bands, sub-6 GHz positioning over \emph{sidelinks}  has recently come into focus, complementary to  enhancements for improved integrity, accuracy, and power efficiency \cite{tr:38859-3gpp22}. 
Sidelink positioning is expected to be an important enabler for several use cases, considering positioning for the \acp{UE} in in-coverage, partial coverage, and out-of-coverage. These include  \ac{V2X},  \ac{IoT} (e.g., in private networks at 3.6--3.7 GHz), and  public safety (e.g., first responders) \cite{tr:38845-3gpp21}. 

Within \ac{V2X}, there have been a number of studies related to sidelink positioning at sub-6 GHz. In terms of overviews, there are several recent papers  \cite{ko2021v2x,bartoletti2021positioning,saily2021positioning,lu2021joint,bazzi2019survey,liu2021highly}. In  
\cite{ko2021v2x},  a broad overview of \ac{V2X} positioning is provided, highlighting the limitations of \ac{TDoA} in terms of synchronization, and proposes to use carrier phase and multipath information.  
In \cite{bartoletti2021positioning}, use cases and corresponding requirements are specified, showing that beyond 5G systems need enhancements in terms of 
deployments, methods, and architectures. 
In \cite{saily2021positioning}, sidelink positioning is advocated as a positioning enabler with lower latency, higher \ac{LoS} probability, and improved coverage,  which should be able to operate both collaboratively with the network and to operate independently when the network is unavailable. Focusing on indoor \ac{IoT} scenarios, \cite{lu2021joint} studies combinations of different measurements together with an \ac{EKF}. The relative merits of \ac{C-V2X} and WiFi-based positioning are discussed and evaluated in 
\cite{bazzi2019survey}, based on the WINNER+ model, indicating a preference of \ac{C-V2X}.  In \cite{liu2021highly}, different positioning systems/architectures  are described for vehicular positioning applications, along with relevant \acp{KPI}. There have also been several studies focusing exclusively on the physical layer \cite{del2017performance,hossain2017high,kakkavas2018multi,JCS_V2X_2022}. \begin{figure}
    \centering
    \includegraphics[width=\columnwidth]{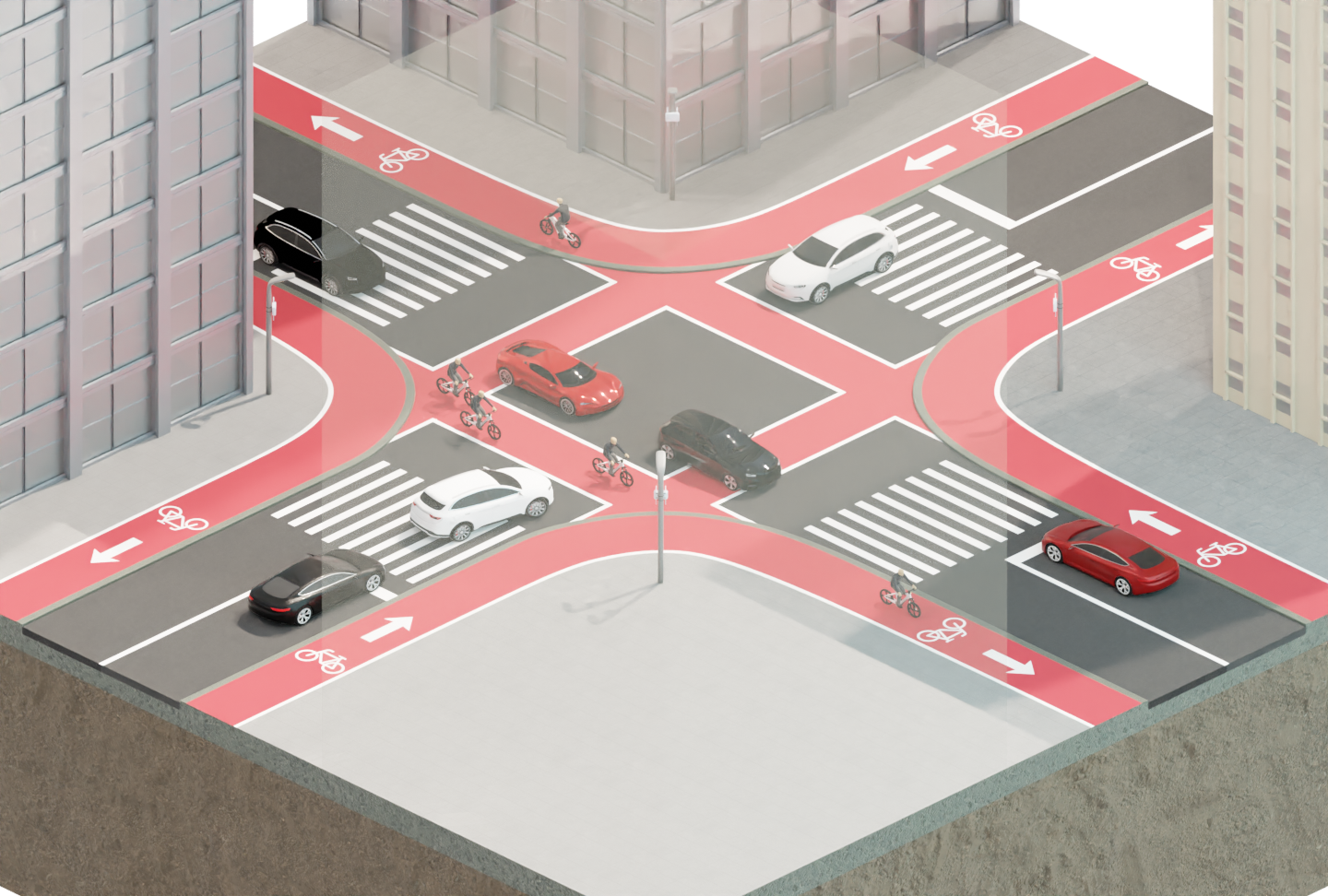}
    \caption{Illustration of an urban traffic situation involving vulnerable road users,  vehicles, and road-side units. }
    \label{fig:city_center}
    \vspace{-5mm}
\end{figure}
Among these, \cite{del2017performance} evaluates \ac{V2I} ranging and \ac{TDoA} positioning for LTE under different statistical channel models. In  \cite{hossain2017high},  \ac{DSRC} positioning is studied with ray-tracing data, combining \ac{AoA} information at \acp{RSU} and \ac{V2V} cooperating ranging links. The use of several arrays per UE is proposed in \cite{kakkavas2018multi} and evaluated in terms of \ac{CRB} at 3.5 and 28 GHz, based on \ac{AoA} and \ac{TDoA} measurements. In 
\cite{JCS_V2X_2022},  sensing with 5G-V2X waveforms is considered, determining the range and Doppler of passive targets via \ac{CRB} analysis under a pure geometric channel, subject to interference. Finally, there are studies that focus mainly on algorithmic aspects, such as \cite{ma2019efficient,liu2021v2x,fouda2021dynamic}. Here, 
\cite{ma2019efficient} proposes a method for V2X localization with a single \ac{RSU} from range measurements over time.
In \cite{liu2021v2x}, \ac{V2V} range and angle and \ac{V2I} \ac{TDoA} measurements are combined to improve positioning. In \cite{fouda2021dynamic}, a dynamic method is proposed to switch between \ac{GNSS} and  NR \ac{V2X} \ac{TDoA} measurements. While the above listed papers adopt widely varying assumption, models, methods, and evaluation methodologies, it is worth pointing out that  \cite{tr:38859-3gpp22} has proposed a common evaluation methodology and a set of common assumptions.

In this paper, we perform a realistic evaluation of sidelink \ac{V2X} \ac{RTT} positioning towards 3GPP Release 18 using ray-tracing data, focusing on operation outside of the network coverage. % . We describe the generic geometric model and relevant channel models, as well as measurements derived from the received signals. In contrast to previous studies, our focus is on operation outside of the network coverage and on geometric channel models, rather than conventional mixed statistical and geometric models. Our specific contributions are: 
\begin{itemize}
    \item \textbf{Use cases:} We describe the relevant V2X use cases towards 3GPP Release 18 and requirements that involve sidelink positioning, as well as the related physical models, and limitations thereof.
    \item \textbf{Novel performance bound:} We propose a novel methodology based on Fisher information analysis to predict positioning performance in the presence of severe multipath, by accounting for inter-path interference. 
    \item \textbf{Methods and evaluation:} 
    We verify the validity of the new performance bound through evaluation of \ac{RTT}-based ranging using ray-tracing data in two use cases: one with \ac{RSU} and one without \ac{RSU}. The simulations show that performance is mainly limited due to multipath induced biases. 
    %We provide simulation results for several scenarios based on Ray-tracing data. 
\end{itemize}
The remainder of this paper is organized as follows. The use cases and system model are introduced in Section \ref{usecase}. The basics of the \ac{FIM} and its three variants are described in Section \ref{FIM}. Ranging and range-based positioning algorithms are presented in Section \ref{range_position}. Simulation results are displayed in
Section \ref{simulation}, followed by our conclusions in Section \ref{conclusion}.

\section{Use Cases and System Model} \label{usecase}
In this section, we describe the different requirement sets and a generic system model.

\subsection{Use Cases and Requirements} \label{use_cases}
In \cite{tr:38845-3gpp21}, three sets of positioning requirements are defined (both for absolute and relative positioning): 
\begin{enumerate}
    \item \emph{Set 1 (low accuracy):} This set requires   10--50 m with 68\%--95\% confidence level, mainly for information provisioning use cases, such as traffic jam warning.
    \item \emph{Set 2 (moderate accuracy):} This set requires   1--3 m with 95\%--99\% confidence level, mainly to support so-called day-1 use cases, including lane change warning (\ac{V2V}), intersection movement assist (\ac{V2V}), and  automated intersection crossing (\ac{V2I}) \cite{5GAA-SysArch}.
    \item \emph{Set 3 (high accuracy):} This set requires 0.1-0.5 m with 95\%--99\% confidence level, to support so-called advanced use cases, such as automated driving or tele-operated driving \cite{5GAA-SysArch}. 
\end{enumerate}
More detailed requirements can be found in \cite[Table 5.1-1]{5GAA-SysArch} and \cite[Table 7.3.2.2-1]{tr:38845-3gpp21}, which also describe the nominal velocity and whether the requirement pertains to absolute or relative positioning.

To exemplify these use cases, Fig.~\ref{fig:city_center} depicts a dense traffic situation in an urban environment. Many road users are trying to cross the intersection. In this situation, \ac{V2X} communication helps to make road traffic safer and more efficient. V2X communication  includes the communication between road users, namely UEs and road infrastructure, i.e., \acp{RSU}. In Fig.~\ref{fig:city_center}, lamp posts are equipped with \acp{RSU}. Within the 3GPP framework \cite{tr:37885-3gpp19}, however, \acp{RSU} are assumed to be mounted at the middle of the intersection. 
Originally, NR-V2X addresses direct communication  between road participants to exchange V2X messages including warnings, information, collective perception etc. Starting with Rel. 18, 3GPP studies the possibility to perform ranging and positioning over sidelink for V2X applications. Especially in difficult outdoor environments where classical positioning techniques, as e.g., GNSS, are blocked or distorted, sidelink positioning arises as a valuable complementing positioning technology. As depicted in Fig.~\ref{fig:city_center}, the street canyons can block the GNSS signals, so that GNSS is considered to be unavailable. %At this intersection, the scenario investigated in this paper covers a VRU crossing use case.

%\todo{@Max: Description of one V2X key use case (UEs can be either traffic participant as cars or bicycles but also Road Side Units (RSU) as communicating traffic lights. In case of RSU a fixed anchor is given.  }

\subsection{System Model}

We consider a scenario with several devices, which may be \acp{UE} or \acp{RSU}. The state components of device $n$, comprising its location (in 2D or 3D) and velocity are denoted by $\boldsymbol{x}_n$ and $\boldsymbol{v}_n$, respectively. For a \ac{RSU}, the state is known and the velocity is $\boldsymbol{v}_n=\boldsymbol{0}$. Devices are not synchronized. %, precluding the use of \ac{TDoA} solutions. Hence our focus is on \ac{RTT}. 
The main functionality is the ability to estimate the \ac{ToA} of the \ac{LoS} path. Our focus is on \ac{OFDM}, where we consider a system with $N_{\text{s}}$ subcarriers with subcarrier spacing $\Delta_f$. 

We drop device indices when possible, so that the received signal at a device, based on the transmission by another device can be expressed as a vector of length $N_{\text{s}}$:\footnote{After appropriate receiver-side processing, such as coarse synchronization, cyclic prefix removal, and FFT \cite{Passive_OFDM_2010}.} 
\begin{align}
    \boldsymbol{y}_t=\sum_{l=0}^{L-1}\alpha_l  ( \boldsymbol{s}_{t} \odot \boldsymbol{a}_{\text{d}}(\tau_l)) e^{\jmath 2 \pi t v_l T_s/\lambda} + \boldsymbol{n}_t, \label{eq:signalModel}
\end{align}
where $t\in \{1,\ldots,T\}$ is the OFDM symbol index, $L$ is the number of  multipath components (which are not necessarily resolvable),  $\alpha_l$ is the complex channel gain  of the $l$-th path,  $\boldsymbol{s}_{t}$ is the vector of pilot signals across the subcarriers of the $t$-th OFDM symbol,  $\boldsymbol{a}_{\text{d}}(\tau_l) \in \mathbb{C}^{N_{\text{s}}}$ is the delay steering vector, as a function of the \ac{ToA} $\tau_l$, with
\begin{align}
    [\boldsymbol{a}_{\text{d}}(\tau_l)]_n = \exp(-\jmath 2 \pi n \tau_l \Delta_f).
\end{align} 
In addition,  $v_l$ is the radial velocity of the $l$-th path, $\lambda$ is the wavelength, and $T_s$ is the OFDM symbol duration (including \ac{CP}). Finally,  
$\boldsymbol{n}_t$ is the \ac{AWGN}, with $\boldsymbol{n}_t \sim \mathcal{CN}(\boldsymbol{0},N_0 \boldsymbol{I})$. The average transmit power is fixed to $P_{\text{tx}}$, so that $\mathbb{E}\{ \Vert\boldsymbol{s}_{t} \Vert^2\}=P_{\text{tx}}/\Delta_f$. 

Under the assumption that the path index $l=0$ correspond to the \ac{LoS} path, the parameter  $\tau_0$ % , $\boldsymbol{d}_{\text{A},0}$ and $\boldsymbol{d}_{\text{D},0}$ 
depends on the geometry through
\begin{align}
    \tau_0& =\Vert \boldsymbol{x}_{\text{rx}}-\boldsymbol{x}_{\text{tx}}\Vert /c + B% \\
    %\boldsymbol{d}_{\text{A},0}& =\boldsymbol{R}^\top(\boldsymbol{o}_{\text{rx}}) \frac{\boldsymbol{x}_{\text{tx}}-\boldsymbol{x}_{\text{rx}}}{\Vert \boldsymbol{x}_{\text{tx}}-\boldsymbol{x}_{\text{rx}}\Vert }\\
    %\boldsymbol{d}_{\text{D},0}& =\boldsymbol{R}^\top(\boldsymbol{o}_{\text{tx}}) \frac{\boldsymbol{x}_{\text{rx}}-\boldsymbol{x}_{\text{tx}}}{\Vert \boldsymbol{x}_{\text{tx}}-\boldsymbol{x}_{\text{rx}}\Vert },
\end{align}
where %, following the rotation conventions in \cite{nazari2022mmwave}, $\boldsymbol{R}(\boldsymbol{o}) \in \text{SO}(3)$ is a rotation matrix that converts from system frame to local frame,
$c$ is the speed of light, and $B$ is a clock bias between the transmitter and the receiver.  In contrast to \ac{TDoA}-based positioning, where the clock bias is removed by using synchronized measurement units, here, we consider an \ac{RTT}-based positioning, where the clock bias can be (approximately) removed following bidirectional transmissions.

\begin{remark}[Impact of velocity]
In \eqref{eq:signalModel}, Doppler can be omitted by considering a coherent transmission interval, which requires that the number of OFDM symbols, say, $T$ is limited to $T \ll  {\lambda\Delta_f}/{v}$, where $v=\max_l v_l$ is the largest (radial) velocity. The velocity also has an impact on the overall tolerable latency \cite{behravan-6g-2022}: if the positioning requirement is $\delta$ meters and the velocity is $v$ meter/second, then the overall latency (including triggering, initialization, transmission, processing, and position reporting) should be completed within approximately $0.1 \delta / v$, in order to have negligible impact on the accuracy.
\end{remark}

Based on the above model, the sidelink positioning problems are as follows. 
%\begin{itemize}
 %   \item 
 \emph{The link-level measurement problem:} From observations of the form \eqref{eq:signalModel}, estimate the \ac{ToA} of the \ac{LoS} path. 
    %\item
    \emph{The relative positioning problem:} Based on the link-level measurements, determine the relative location of two devices. 
    %\item 
    \emph{The absolute positioning problem:} Based on the link-level measurements of one \ac{UE} with respect to several \acp{RSU}, determine the location of the \ac{UE} in a global coordinate frame. 
%\end{itemize}

%38.845 (maybe summarize Table Table 7.3.2.2-1??)
%Set 1: 10 – 50 m with 68 – 95 \% confidence level. 
%-	Set 2: 1 – 3 m with 95 – 99 \% confidence level. 
%-	Set 3: 0.1 – 0.5 m with 95 – 99 \% confidence %level.
%For V2X use case, the ITS-dedicated spectrum can be considered for PC5 interface, and the spectrum licensed to mobile network operators (including FR2) and the unlicensed spectrum can be considered for both Uu and PC5 interfaces. Note that there is no mechanism corresponding to regulatory requirements to use unlicensed spectrum in Rel-17 NR sidelink. 

%\begin{itemize}
    %\item Generic system model, signal model, channel model.
    %\item Channel model: LOS + dense multipath (statistical modeling) or many paths with deterministic modeling
    %\item Synchronization assumptions
    %\item Listing of key use cases (V2X, public safety, commercial and IIOT). 
    %\item Requirements? Can be taken from 3GPP. Different sets were defined. E.g. Set 2: 1 – 3 m with 95 – 99 confidence level
%    \item Description of one V2X key use case (UEs can be either traffic participant as cars or bicycles but also Road Side Units (RSU) as communicating traffic lights. In case of RSU a fixed anchor is given.  
%\end{itemize}

\section{Fisher Information Analysis} \label{FIM}
Fisher information theory \cite{VanTrees} is a common approach to predict the performance of the measurement and positioning sub-problems, and also serves as a design tool for, e.g., waveform optimization or node placement \cite{shen_localization}. 

\subsection{Fundamentals of \ac{FIM}}
Given an abstract observation model $ \boldsymbol{r}=\boldsymbol{\mu}(\boldsymbol{\eta},\boldsymbol{\kappa}) + \boldsymbol{n}$, 
where $\boldsymbol{\eta}$ are the geometric parameters of interest (e.g., delay of the \ac{LoS} path, but in general also the \ac{AoA}, \ac{AoD}, and Doppler), $\boldsymbol{\kappa}$ are nuisance parameters (e.g., channel gains, as well as delays of the \ac{NLoS} paths), $\boldsymbol{\mu}(\cdot)$ is a known possibly non-linear mapping, and $\boldsymbol{n}$ is complex white noise with variance $N_0$. We also introduce $\boldsymbol{\xi}=[\boldsymbol{\eta}^\top,\boldsymbol{\kappa}^\top]^\top$.\footnote{For convenience, we impose that $\boldsymbol{\xi}$ is real, so that any complex variable should be broken down into real and imaginary parts, or amplitude and phase.}  The FIM of $\boldsymbol{\xi}$ is given by
\begin{align}
    \boldsymbol{J}(\boldsymbol{\xi})=\frac{2}{N_0} \Re\Big \{ \big(\frac{\partial \boldsymbol{\mu}(\boldsymbol{\xi})}{\partial \boldsymbol{\xi}}\big)^{\mathsf{H}}
    \frac{\partial \boldsymbol{\mu}(\boldsymbol{\xi})}{\partial \boldsymbol{\xi}}\Big\},
\end{align}
where ${\partial \boldsymbol{\mu}(\boldsymbol{\xi})}/{\partial \boldsymbol{\xi}} \in \mathbb{C}^{\text{dim}(\boldsymbol{\mu}) \times \text{dim}(\boldsymbol{\xi})}$ is the gradient matrix. Here,  $\text{dim}(\cdot)$ returns the dimension of the argument. The FIM is a positive semi-definite matrix with the following property (some technical conditions apply):
\begin{align}
    \mathbb{E} \big \{ (\boldsymbol{\xi}-\hat{\boldsymbol{\xi}})((\boldsymbol{\xi}-\hat{\boldsymbol{\xi}}))^\top \big\} \succeq  \boldsymbol{J}^{-1}(\boldsymbol{\xi}) + \boldsymbol{b}\boldsymbol{b}^\top \doteq \boldsymbol{\Sigma}(\boldsymbol{\xi})\label{eq:FIMCRB}
\end{align}
for any estimator, where $\boldsymbol{b}$ denotes the bias of the estimator. 
%for any unbiased estimator (some technical conditions apply). In case there is a fixed bias $\boldsymbol{b}$, then the right-hand-side of \eqref{eq:FIMCRB} should be modified to $\boldsymbol{J}^{-1}(\boldsymbol{\xi})+\boldsymbol{b}\boldsymbol{b}^\top$ in order to account for this bias. 
To compute a lower bound on the error covariance of any estimator of $\boldsymbol{\eta}$, we simply take the corresponding sub-matrix of $\boldsymbol{\Sigma}(\boldsymbol{\xi})$, i.e., $\boldsymbol{\Sigma}(\boldsymbol{\eta})=[\boldsymbol{\Sigma}(\boldsymbol{\xi})]_{1:\text{dim}(\boldsymbol{\eta}),1:\text{dim}(\boldsymbol{\eta})}$.

%$ \boldsymbol{J}^{-1}(\boldsymbol{\xi})$ to obtain $ \boldsymbol{J}^{-1}({\boldsymbol{\eta}})$, or we can obtain $ \boldsymbol{J}({\boldsymbol{\eta}})$, i.e., the equivalent \ac{FIM} of $\boldsymbol{\eta}$ under unknown nuisance parameters, by the Schur's complement formula to the original larger $\boldsymbol{J}(\boldsymbol{\xi})$. 
\subsection{Three Variants of \ac{FIM}}
In the literature, there are several ways that the \ac{FIM} of the channel parameters related to the observation \eqref{eq:signalModel} can be computed. We recall that we are interested only in the \ac{LoS} \ac{ToA}, so that   $\boldsymbol{\eta}=[\tau_0]$. 

\subsubsection{\ac{LoS}-only bound} In this approach, only the term for $l=0$ is retained in \eqref{eq:signalModel}. This is equivalent to having perfect knowledge of the \ac{NLoS} parameters. The corresponding lower bound on the error covariance is denoted by $\boldsymbol{\Sigma}_{\text{LoS}}(\tau_0)$, and typically does not require any bias term. 
%\ac{FIM}, which we denote by $\boldsymbol{J}_{\text{LoS}}({\boldsymbol{\eta}})$, 
It has the property that $ \mathbb{E} \{ \vert \tau_0-\hat{\tau}_0\vert^2 \} \gg \boldsymbol{\Sigma}_{\text{LoS}}(\tau_0)$. However, due to the overly optimistic assumption on exact knowledge of the \ac{NLoS} parameters, this bound is likely very loose (i.e., $\geq$ is likely $\gg$) and thus not useful.

\subsubsection{All-paths bound} In
this approach, all the multipath components are retained, so that the delays and angles for all $L$ paths are estimated. Since paths outside the \ac{LoS} path's \emph{resolution cell}\footnote{The $l$-th path ($l>0$) is in the \ac{LoS} resolution cell when $|\tau_l-\tau_0|\le \beta / (N_{\text{s}}\Delta_f)$, where $1<\beta<2$.} do not affect the estimation accuracy of $\tau_0$, such paths can be removed in order to tighten the bound. 
We denote the corresponding error covariance bound by $\boldsymbol{\Sigma}_{\text{all}}(\tau_0)$, again without any bias term. 
%equivalent \ac{FIM} by $\boldsymbol{J}_{\text{NLoS}}({\boldsymbol{\eta}})$. 
Since the paths in the \ac{LoS} resolution cell are by definition not  resolvable, it is possible that the FIM is (nearly) rank-deficient. In that case, $ \mathbb{E} \{ \vert \tau_0-\hat{\tau}_0\vert^2 \} < \boldsymbol{\Sigma}_{\text{all}}(\tau_0)$, so that the inequality can occur in the opposite direction (i.e., $\boldsymbol{\Sigma}_{\text{all}}(\tau_0)$ is no longer a lower bound). 

\subsubsection{Weighted average approximation (WAA)}
To address the shortcomings of the \ac{LoS}-only bound (which is too loose) and the all-paths bound (which is not a bound), a pragmatic solution is proposed. While strictly speaking it is not a bound, it serves as a useful approximation of the performance of real estimators.  We collect the indices of all the path in the resolution cell of the \ac{LoS} path into the set 
%, which provides an approximation of the error covariance that is neither too loose from below or above can be found by merging all the paths within the resolution cell\footnote{A resolution cell is the interval of angles and delays that are not resolvable with the \ac{LoS} path. Note that resolution cells maybe be large in one domain (e.g., bandwidth when bandwidth is limited) and large in another (e.g., \ac{AoA}, if the number of receive antennas is large)} of the \ac{LoS} path. Collecting the corresponding paths in a set 
$\mathcal{L}\subseteq \{0,1,\ldots,L-1\}$ (always $0\in \mathcal{L}$). We then introduce the weight of each path 
$w_l={|\alpha_l|}/{\sum_{l' \in \mathcal{L}}|\alpha_{l'}|}$, such that $\sum_{l \in \mathcal{L}} w_l=1$. The merged \ac{ToA} is computed as 
\begin{align}\label{eq_merged_toa}
    \bar{\tau}_0 = \sum_{l \in \mathcal{L}} \tau_l w_l.
\end{align}
Note that paths with a comparably large amplitude have a higher impact on the merged \ac{ToA} in \eqref{eq_merged_toa}. We determine the merged 
     complex channel gain as $ \bar{\alpha}_0=\sum_{l \in \mathcal{L}} \alpha_l $. 
%we determine the merged 
   %  complex channel gain is computed as $ \bar{\alpha}_0=\sum_{l \in \mathcal{L}} \alpha_l $, while the merged  and introduce $w_l={|\alpha_l|}/{\sum_{l' \in \mathcal{L}}|\alpha_{l'}|}$. Then, the merged 
    %\ac{ToA} is given by
    %\begin{align}
    %    \bar{\tau}_0& =\sum_{l \in \mathcal{L}} \tau_l w_l.%\\
       %\bar{\boldsymbol{d}}_{\text{A},0} & =\mathcal{P}\big(\sum_{l \in \mathcal{L}} \boldsymbol{d}_{\text{A},l}w_l\big)\\
       %\bar{\boldsymbol{d}}_{\text{D},0} & =\mathcal{P}\big(\sum_{l \in \mathcal{L}} \boldsymbol{d}_{\text{D},l} w_l\big),
    %\end{align}
%    where $\mathcal{P}(\cdot)$ is a suitable project operator onto the set of $3 \times 1$ unit norm vectors. 
    We then compute the \ac{FIM} of 
    $[\bar{\tau}_0,\bar{\alpha}_0]^\top$, which comprises the parameters of the actual (effective) path seen by the estimator. Finally, the expression for the novel WAA is     \begin{align}
        \mathbb{E} \{ \vert \tau_0-\hat{\tau}_0\vert^2 \} \approx  
        [\boldsymbol{J}^{-1}([\bar{\tau}_0,\bar{\alpha}_0]^\top)]_{1,1}+(\tau_0-\bar{\tau}_0)^2. 
    \end{align}
    %the merged \ac{LoS} parameters only, as well as the corresponding bias (e.g., $b_{\tau}=(\tau_0-\bar{\tau}_0)$) to end up with an approximation of the error covariance. While the resulting value is neither an upper or a lower bound, it can account for path non-resolvability. 
    In the special case where the $L$ paths are resolvable, $w_0=1$ and the bias will be zero, thus reverting  to the LoS-only bound. 

\section{Algorithms for Ranging and  Positioning} \label{range_position}
In this section, we provide a recap of standard approaches for distance estimation and for positioning, as well as their main error sources.

\subsection{\ac{ToA} Estimation and RTT} \label{TOAest}
% \todo{TOA estimation methods. Time-domain MUSIC works well under simulated data and moves well beyond standard bandwidth limitations. Discussion of the error sources (noise, multipath). }

For OFDM waveform, TOA estimation can be performed via IDFT over subcarriers in \eqref{eq:signalModel}, after removing the effect of data symbols \cite{OFDM_Radar_Corr_TAES_2020,RadCom_Proc_IEEE_2011,OFDM_Radar_Phd_2014,Passive_OFDM_2010}, i.e.,
\begin{align}\label{eq_yy_fft}
   \boldsymbol{z}= \left| \FF^{H}_{N_s} \sum_{t=1}^{T}  \boldsymbol{w} \odot (\boldsymbol{y}_t \oslash \boldsymbol{s}_t ) \right|^2   \in \complexset{N_s}{1} ~,
\end{align}
where $\FF_N \in \complexset{N}{N}$ represents a unitary DFT matrix, $\boldsymbol{w} \in \realset{N_s}{1}$ is the window (e.g., a Hamming window) to suppress side-lobes,  and $\oslash$ denotes element-wise division. Note that the steering vector $\boldsymbol{a}_{\text{d}}(\tau)$ in \eqref{eq:signalModel} corresponds to DFT matrix columns, meaning that $ \boldsymbol{z}$ in \eqref{eq_yy_fft} corresponds to the delay spectrum, where peaks indicate delays for different paths.
%Using coherent processing, one can jointly estimate angles and delays of multiple paths via
% \begin{align}\label{eq_yy_fft1}
%     \lvert \FF^{H}_{N_{rx}} \boldsymbol{Y}_t \FF^{*}_{N_s} \rvert^2 \in \complexset{N_{rx}}{N_s} ~,
% \end{align}
Multipath resolvability in delay domain, which is a function of the available bandwidth, can significantly affect the performance of \ac{ToA} estimation in \eqref{eq_yy_fft} \cite{shen_localization,toa_multipath}. Unresolvable \ac{NLoS} paths in \eqref{eq:signalModel} can lead to biases in \ac{ToA} estimation, while the level of noise power determines the variance of \ac{ToA} estimates \cite{shen_localization}.

%\subsection{From \ac{ToA} to \ac{TDoA} and \ac{RTT}}
% \todo{a brief description on how tdoa and rtt are computed and what the error sources are (clock errors) .}

\ac{ToA} obtained via \eqref{eq_yy_fft} can be converted to distance only under perfect synchronization between the transmitter and the receiver. In the absence of synchronization, delay measurements involve the combined impact of distance and relative clock offset. 
% between the TX and RX. A remedy to overcome this issue is to resort to TDOA measurements using signals either from a single anchor arriving through multiple paths, or from the resolvable LOS paths associated to multiple anchors \cite{ko2021v2x}. In the latter case, anchors (e.g., RSUs) must be perfectly synchronized among themselves \cite{bartoletti2021positioning}. TDOA can be computed by taking the difference of multiple delay/TOA measurements with respect to a chosen reference measurement. This enables cancelling out the clock offset between the anchor(s) (TX) and the RX \cite{ko2021v2x}. 
RTT offers a solution to deal with the clock offset in range-based positioning \cite{5g_nr_rel16,5g_nr_v2x_driving}. In multi-cell (or, multi-anchor) RTT, the user can measure the round-trip time to multiple anchors and infer distances by subtracting the known processing times at each anchor. Since two noisy measurements are combined to obtain a single range estimate, the error variance in RTT is double the value in a single link, under perfect knowledge of the processing times.

\subsection{\ac{RTT}-based Positioning}
% \todo{Description on how to compute the position and listing of error sources (geometry, errors in references)}

Given RTT estimates, converted to distances, of the form $\hat{d}_i=\Vert \boldsymbol{x}-\boldsymbol{x}_i\Vert +\omega_i$, where $\boldsymbol{x}$ is the unknown location of the device to be localized,  $\boldsymbol{x}_i$ is the location of the $i$-th reference node, and $\omega_i \sim \mathcal{N}(0,\sigma^2_i)$ is the measurement noise (here modeled as a Gaussian random variable, 
%Based on the distance estimates $\hat{d}_i$, 
the ML estimate of the unknown location can be obtained via 
\begin{align} \label{eq_ml}
    \widehat{\boldsymbol{x}} = \arg \min_{\boldsymbol{x}} \, \sum_{i=1}^{I} \frac{1}{\sigma^2_i} (\hat{d}_i - \Vert \boldsymbol{x}-\boldsymbol{x}_i\Vert)^2 ~.
\end{align}
Solving \eqref{eq_ml} via gradient descent requires a good initial point, which can be obtained in %via LS approaches in 
closed-form \cite{zhu2009simple}.

% \todo{should we be more general?})
% \todo{add ML optimization} \todo{add initial estimate from LS intersection of circles, maybe from \cite{zhu2009simple}?}

% \todo{Can we also add the LS initializer based on the squared distances?}

\section{Simulations} \label{simulation}

\subsection{Scenarios}
In this paper, we will follow the 3GPP \ac{RSU} deployment procedure according to \cite{tr:37885-3gpp19}.
We define two VRU crossing use cases as follows. 
\emph{Scenario 1}: Let us assume a bicycle is driving on the bicycle lane aiming to cross the intersection. At the same time, another vehicle wants to cross the intersection as well. Due to buildings, the line of sight between the vehicle and the bicycle is blocked or at least blocked for a considerable amount of time. Nevertheless, the car and the bicycle are communicating to the RSU. To avoid a collision at the intersection, the aim is to determine the relative position between the vehicle and the bicycle and either send a warning or initiate an emergency brake. 
\emph{Scenario 2} considers the same setting without the presence of the RSU. It is assumed that the bicycle moves with a speed of 15 km/h whereas the vehicle drives at a speed of 50 km/h  \cite{tr:37885-3gpp19}.

\begin{figure}
    \centering
    \includegraphics[width=0.5\textwidth]{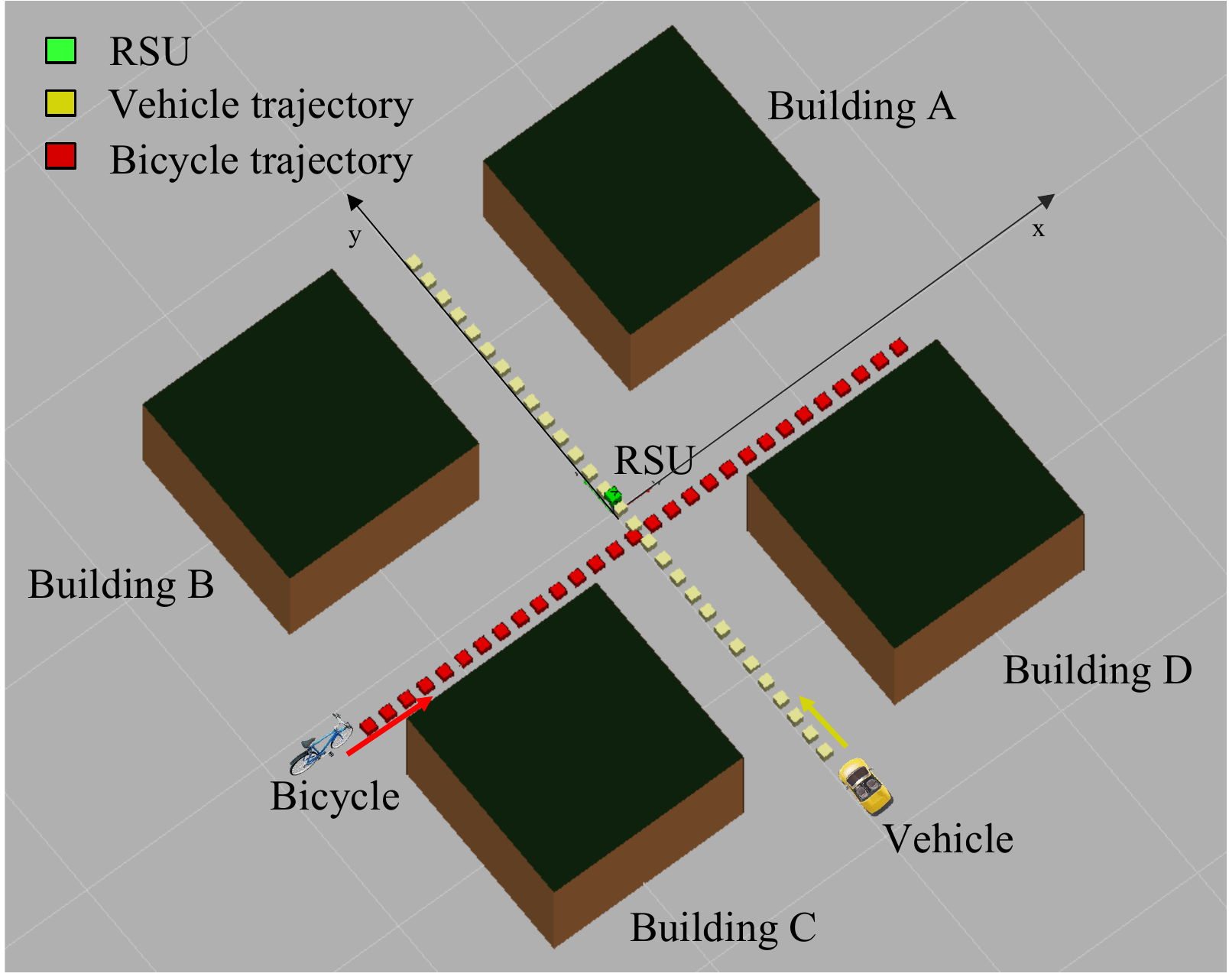}
    \caption{The ray-tracing simulation environment with a single \ac{RSU} at the center of the intersection with location $[0 \, \text{m},0 \, \text{m},10\, \text{m}]^{\mathsf{T}}$, a vehicle moves from down to up (alongside the line $x=1.6\, \text{m}$), a bicycle moves from left to right (alongside the line $y=-7\, \text{m}$), and four buildings.}
    \label{fig:raytracing_env}\vspace{-5mm}
\end{figure}
\subsection{Simulation Environment}

%We consider two urban sidelink scenarios as described in Section \ref{use_cases}. 
Both scenarios are simulated using the REMCOM Wireless InSite\textregistered  ray-tracer \cite{WirelessInSite}. 
%The first scenario has a single \ac{RSU}, two \acp{UE} and 4 buildings as illustrated in Fig.~\ref{fig:raytracing_env}. 
In \emph{Scenario 1}, illustrated in Fig.~\ref{fig:raytracing_env}, the \ac{RSU} is located at the center of the intersection with location $[0 \, \text{m},0 \, \text{m},10\, \text{m}]^{\mathsf{T}}$.  Two \acp{UE} are a vehicle and a bicycle. The vehicle moves alongside the vehicle lane (with its antenna at the height of $1.5\, \text{m}$), where the lane is from $[1.6 \, \text{m},-70 \, \text{m},0 \, \text{m}]^{\mathsf{T}}$  to $[1.6 \, \text{m},70 \, \text{m},0 \, \text{m}]^{\mathsf{T}}$. The bicycle moves alongside the bicycle lane (with its antenna at the height of $1\, \text{m}$), where the lane is from $[-70 \, \text{m},-7 \, \text{m},0 \, \text{m}]^{\mathsf{T}}$  to $[70 \, \text{m},-7 \, \text{m},0 \, \text{m}]^{\mathsf{T}}$. 
The four buildings are $50 \, \text{m}$ long, $50 \, \text{m}$ wide, and $30 \, \text{m}$ high, with their centers located at $[\pm 45 \, \text{m},\pm 45 \, \text{m},15\, \text{m}]^{\mathsf{T}}$, respectively. All the \ac{RSU} and the \acp{UE} are equipped with a single omnidirectional antenna. Every transmission interval, the \ac{RSU} sends signals to the \acp{UE} via the propagation environment, but there is no communication between two \acp{UE}. In  \emph{Scenario 2},  there is no \ac{RSU}, so that the vehicle sends signals to the bicycle every time step. The vehicle starts at $[1.6 \, \text{m},-70 \, \text{m},1.5\, \text{m}]^{\mathsf{T}}$, and moves to the end of the vehicle lane at $[1.6 \, \text{m},70 \, \text{m},1.5\, \text{m}]^{\mathsf{T}}$ with a speed of $14\, \text{m/s}$, and the bicycle starts at $[-16.4 \, \text{m},-7 \, \text{m},1\, \text{m}]^{\mathsf{T}}$ alongside the bicycle lane with a speed of $4.0\, \text{m/s}$. Then, the collision can occur after $4.5\, \text{s}$. 

In terms of signal parameters, the  carrier frequency is $5.9~\text{GHz}$. We consider the OFDM pilot signals with $12$ symbols with a constant amplitude, and $20~\text{MHz}$ total bandwidth with  $167$  subcarriers and $120~\text{kHz}$ subcarrier spacing, corresponding to a pilot duration of $100.2~\text{us}$. The transmitted power is $10~\text{dBm}$, the noise spectral density is $-174~\text{dBm/Hz}$, and the receiver noise figure is $8~\text{dB}$. RTT measurements are generated every 100 ms. 
%The ground-truth channels are generated by Ray-tracing simulation. 

For both scenarios, we evaluate the ranging performance by the \ac{RMSE} over 100 Monte Carlo simulations, where each estimate can be acquired by firstly generating signals via \eqref{eq:signalModel} with Ray-tracing data,  applying the \ac{ToA} estimation  in Section \ref{TOAest} and converting the estimates into distances. We also compare the \acp{RMSE} with three \acp{REB}, obtained by taking the square root of the bound on the error variance of the \ac{ToA} and multiplying with the speed of light. This leads to  the LoS-only REB, the all-paths REB, and the proposed WAA REB.  All codes were written in MATLAB R2018b, and  simulations and experiments were run on a Lenovo ThinkPad T480s with a 1.8 GHz 4-Core Intel Core i7-8850U processor and 24~Gb~memory. 

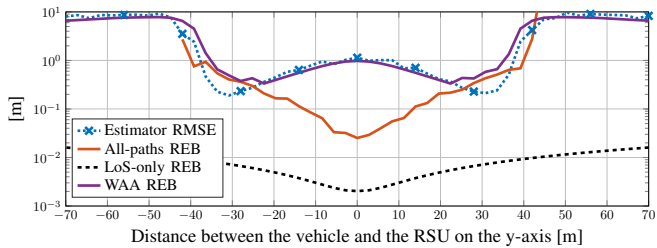
\begin{figure}
\center
\definecolor{mycolor1}{rgb}{0.00000,0.44700,0.74100}%
\definecolor{mycolor2}{rgb}{0.85000,0.32500,0.09800}%
\definecolor{mycolor3}{rgb}{0,0,0}%
\definecolor{mycolor4}{rgb}{0.49400,0.18400,0.55600}%
\begin{tikzpicture}[scale=0.8\linewidth/14cm]

\begin{axis}[%
width=6.028in,
height=2.009in,
scale only axis,
unbounded coords=jump,
xmin=-70,
xmax=70,
xlabel style={font=\color{white!15!black},font=\Large},
xlabel={Distance between the vehicle and the RSU on the y-axis [m]},
ymode=log,
ymin=0.001,
ymax=10,
ylabel style={font=\color{white!15!black},font=\Large},
ylabel={[m]},
yminorticks=true,
%axis x line*=bottom,
%axis y line*=left,
xmajorgrids,
ymajorgrids,
axis background/.style={fill=white},
legend style={at={(0.27,0.45)},legend cell align=left, align=left, draw=white!15!black,font=\large}
]
\addplot [color=mycolor1, dotted, line width=2.0pt,mark=x, mark repeat=5,mark options={solid},mark size=4pt]
  table[row sep=crcr]{%
70	8.2970962273073\\
67.2	7.38237628692777\\
64.39999	8.50050879495954\\
61.60001	8.64693999707331\\
58.8	8.78609270352784\\
56	9.05740364009833\\
53.2	8.3674418095311\\
50.40001	9.6770473496655\\
47.60001	7.45558205563486\\
44.8	7.03205788740526\\
42	4.21880898720673\\
39.20001	2.44631137462686\\
36.40001	0.509803770837992\\
33.60001	0.241856163718767\\
30.8	0.211920259107629\\
28	0.228254607431809\\
25.2	0.272669937469318\\
22.39999	0.368389663631076\\
19.59999	0.412543690637496\\
16.79999	0.536622224974401\\
13.99998	0.705078977345553\\
11.19998	0.689180467418059\\
8.399977	1.01084927158879\\
5.599974	1.01575703655749\\
2.799971	0.985763691334145\\
-3.213741e-05	1.13826492372548\\
-2.800035	1.02738595834176\\
-5.600038	0.835828045727544\\
-8.400042	0.9517788897052\\
-11.20004	0.784037530255835\\
-14.00005	0.637180865491065\\
-16.80005	0.583387215497089\\
-19.60005	0.450042882009324\\
-22.40006	0.36100419740616\\
-25.20006	0.273921032609682\\
-28.00006	0.233440354505001\\
-30.80007	0.189957825337008\\
-33.60007	0.226597159827525\\
-36.40007	0.464693989546502\\
-39.20007	2.42408304348441\\
-42.00008	3.53489385542999\\
-44.80008	7.7198552516744\\
-47.60008	9.21442789723343\\
-50.40009	8.40774373061713\\
-53.20009	8.77964212325155\\
-56.00009	8.65462047729212\\
-58.80009	8.47472500159768\\
-61.60008	8.63276281460891\\
-64.40007	7.42522855411169\\
-67.20006	8.59506699010332\\
-70.00005	8.02340025849856\\
};
\addlegendentry{Estimator RMSE}

\addplot [color=mycolor2, line width=2.0pt]
  table[row sep=crcr]{%
70	inf\\
67.2	inf\\
64.39999	inf\\
61.60001	inf\\
58.8	inf\\
56	inf\\
53.2	inf\\
50.40001	inf\\
47.60001	inf\\
44.8	61.6220960180601\\
42	2.75509318096641\\
39.20001	0.687194780753915\\
36.40001	0.639251414178629\\
33.60001	0.510895394460789\\
30.8	0.429965522931822\\
28	0.341530630141623\\
25.2	0.249795972611999\\
22.39999	0.216576454101637\\
19.59999	0.20753084328548\\
16.79999	0.132563944031606\\
13.99998	0.112213555669774\\
11.19998	0.0649543513786708\\
8.399977	0.0555738097025537\\
5.599974	0.0398321774671638\\
2.799971	0.0291623404736184\\
-3.213741e-05	0.0252203679531004\\
-2.800035	0.0319667326189675\\
-5.600038	0.0332369559816384\\
-8.400042	0.0642929196092308\\
-11.20004	0.0848032881314534\\
-14.00005	0.113769004205595\\
-16.80005	0.163604439622623\\
-19.60005	0.165760248638464\\
-22.40006	0.209025542297017\\
-25.20006	0.298947889399994\\
-28.00006	0.362826646102247\\
-30.80007	0.40480533143343\\
-33.60007	0.542084203490707\\
-36.40007	0.937039317910486\\
-39.20007	0.755278725572318\\
-42.00008	2.72444507100944\\
-44.80008	inf\\
-47.60008	inf\\
-50.40009	inf\\
-53.20009	inf\\
-56.00009	inf\\
-58.80009	inf\\
-61.60008	inf\\
-64.40007	inf\\
-67.20006	inf\\
-70.00005	inf\\
};
\addlegendentry{All-paths REB}

\addplot [color=mycolor3, dashed, line width=2.0pt]
  table[row sep=crcr]{%
70	0.0160445329923981\\
67.2	0.0154126416362986\\
64.39999	0.0147817917873996\\
61.60001	0.014150672524616\\
58.8	0.0135215687490023\\
56	0.0128907171935636\\
53.2	0.012262445229979\\
50.40001	0.0116339469904224\\
47.60001	0.0110072056513633\\
44.8	0.0103819057192902\\
42	0.00975724209184846\\
39.20001	0.00913433805797294\\
36.40001	0.00851387129505732\\
33.60001	0.00789636255916457\\
30.8	0.0072816044005457\\
28	0.00667078710578359\\
25.2	0.00606634465247407\\
22.39999	0.00546861161601763\\
19.59999	0.00488064343549899\\
16.79999	0.00430702190583903\\
13.99998	0.00375385143442729\\
11.19998	0.0032320426316711\\
8.399977	0.00275979663401044\\
5.599974	0.00236852013263941\\
2.799971	0.00210803142445443\\
-3.213741e-05	0.00202991139086203\\
-2.800035	0.00210803142445444\\
-5.600038	0.00236852013263941\\
-8.400042	0.00276011438563079\\
-11.20004	0.0032320426316711\\
-14.00005	0.00375385143442729\\
-16.80005	0.00430702190583903\\
-19.60005	0.00488064343549899\\
-22.40006	0.00546861161601763\\
-25.20006	0.00606634465247407\\
-28.00006	0.00667078710578359\\
-30.80007	0.0072816044005457\\
-33.60007	0.00789636255916457\\
-36.40007	0.00851387129505732\\
-39.20007	0.00913433805797294\\
-42.00008	0.00975724209184846\\
-44.80008	0.0103819057192902\\
-47.60008	0.0110072056513633\\
-50.40009	0.0116339469904224\\
-53.20009	0.012262445229979\\
-56.00009	0.0128907171935636\\
-58.80009	0.0135215687490023\\
-61.60008	0.014150672524616\\
-64.40007	0.0147817917873995\\
-67.20006	0.0154126416362986\\
-70.00005	0.0160445329923981\\
};
\addlegendentry{LoS-only REB}

\addplot [color=mycolor4, line width=2.0pt]
  table[row sep=crcr]{%
% 70	6.58091639105836\\
% 67.2	6.78500205552436\\
% 64.39999	6.99770299231624\\
% 61.60001	7.2173742517603\\
% 58.8	7.44766508470461\\
% 56	7.61465317508626\\
% 53.2	7.70069139564691\\
% 50.40001	7.57330876855277\\
% 47.60001	5.07167630428988\\
% 44.8	5.02525009302342\\
% 42	4.72703647252908\\
% 39.20001	0.110471364715516\\
% 36.40001	0.135769051606742\\
% 33.60001	0.164767866921381\\
% 30.8	0.197948206979879\\
% 28	0.236445392025847\\
% 25.2	0.281315162387697\\
% 22.39999	0.334081444657494\\
% 19.59999	0.396326036515321\\
% 16.79999	0.469959092945072\\
% 13.99998	0.556565229492818\\
% 11.19998	0.6560686543619\\
% 8.399977	0.764388480351211\\
% 5.599974	0.869082555178748\\
% 2.799971	0.947789587745617\\
% -3.213741e-05	0.977234273199649\\
% -2.800035	0.947787252433303\\
% -5.600038	0.869080392514773\\
% -8.400042	0.764443606885588\\
% -11.20004	0.656067035130158\\
% -14.00005	0.556561984037272\\
% -16.80005	0.469960157123149\\
% -19.60005	0.39633156304522\\
% -22.40006	0.33408059067328\\
% -25.20006	0.281313817500737\\
% -28.00006	0.236433159342317\\
% -30.80007	0.197937636642903\\
% -33.60007	0.164757191390072\\
% -36.40007	0.135770511352875\\
% -39.20007	0.110477550544671\\
% -42.00008	4.64006944409932\\
% -44.80008	5.07097041120031\\
% -47.60008	5.09788602341824\\
% -50.40009	7.57871877261707\\
% -53.20009	7.6939067002491\\
% -56.00009	7.60940247272111\\
% -58.80009	7.44115689349155\\
% -61.60008	7.21197220373898\\
% -64.40007	6.99323732060499\\
% -67.20006	6.78166200771345\\
% -70.00005	6.57917019180968\\
% };
70	6.58911093910717\\
67.2	6.78495001740982\\
64.39999	6.99770642836808\\
61.60001	7.21740276942695\\
58.8	7.4477554416135\\
56	7.61477394891773\\
53.2	7.70614589087606\\
50.40001	7.78585060819267\\
47.60001	7.78605637529772\\
44.8	7.59876868460608\\
42	6.57212001213513\\
39.20001	4.51536085192517\\
36.40001	1.34206934458743\\
33.60001	0.660953498858174\\
30.8	0.580774557831599\\
28	0.425265148187592\\
25.2	0.435209716227043\\
22.39999	0.334079191454086\\
19.59999	0.396325941449528\\
16.79999	0.469959900822744\\
13.99998	0.556563021598003\\
11.19998	0.656070797221403\\
8.399977	0.764387458013131\\
5.599974	0.869082450097599\\
2.799971	0.947789565949275\\
-3.213741e-05	0.97723408621454\\
-2.800035	0.947787906157684\\
-5.600038	0.869080793085168\\
-8.400042	0.764443604617919\\
-11.20004	0.656067380151497\\
-14.00005	0.556563496461615\\
-16.80005	0.469960332201649\\
-19.60005	0.396327846909243\\
-22.40006	0.334082565686671\\
-25.20006	0.429960725341358\\
-28.00006	0.376622522143438\\
-30.80007	0.491711080076637\\
-33.60007	0.653133256851049\\
-36.40007	1.41202164970694\\
-39.20007	4.50230915744055\\
-42.00008	6.52637429413183\\
-44.80008	7.58325329502619\\
-47.60008	7.80712765780128\\
-50.40009	7.77255053148785\\
-53.20009	7.69825569806351\\
-56.00009	7.60942017243743\\
-58.80009	7.4411654206505\\
-61.60008	7.21197266928212\\
-64.40007	6.99314871750823\\
-67.20006	6.78165834084339\\
-70.00005	6.58737050849975\\
};
\addlegendentry{WAA REB}

\end{axis}
\end{tikzpicture}%
\caption{Scenario 1 (vehicle): Comparison among the \ac{RMSE} of estimated range between the vehicle and the \ac{RSU} and three \acp{REB}.}
\label{car}
%\vspace{-4mm} 
\end{figure}

\begin{figure}
\center
\definecolor{mycolor1}{rgb}{0.00000,0.44700,0.74100}%
\definecolor{mycolor2}{rgb}{0.85000,0.32500,0.09800}%
\definecolor{mycolor3}{rgb}{0,0,0}%
\definecolor{mycolor4}{rgb}{0.49400,0.18400,0.55600}%
\begin{tikzpicture}[scale=0.8\linewidth/14cm]
\begin{axis}[%
width=6.028in,
height=2.009in,
at={(1.011in,2.014in)},
scale only axis,
unbounded coords=jump,
xmin=-70,
xmax=70,
xlabel style={font=\color{white!15!black},font=\Large},
xlabel={Distance between the bicycle and the RSU on the
x-axis [m]},
ymode=log,
ymin=0.001,
ymax=10,
ylabel style={font=\color{white!15!black},font=\Large},
ylabel={[m]},
yminorticks=true,
%axis x line*=bottom,
%axis y line*=left,
xmajorgrids,
ymajorgrids,
axis background/.style={fill=white},
legend style={at={(0.27,0.45)},legend cell align=left, align=left, draw=white!15!black,font=\large}
]

\addplot [color=mycolor1, dotted, line width=2.0pt,mark=x, mark repeat=5,mark options={solid},mark size=4pt]
  table[row sep=crcr]{%
-70	7.60657039853723\\
-67.2	8.10983288257608\\
-64.4	7.01902019191861\\
-61.60001	8.02349854875935\\
-58.8	7.66206935989721\\
-56	8.0597092419417\\
-53.20001	8.14169427319387\\
-50.40001	8.29970218017977\\
-47.60001	5.91387136699751\\
-44.80002	3.71559199595186\\
-42.00002	3.50959056493522\\
-39.20003	3.30286779771932\\
-36.40002	3.52973198051642\\
-33.60001	2.4009096373607\\
-30.8	0.269467365257598\\
-27.99998	0.145948588160898\\
-25.19998	0.1746053526313\\
-22.39996	0.220570330720349\\
-19.59996	0.243138948668272\\
-16.79994	0.295992775987283\\
-13.99994	0.380772621973213\\
-11.19992	0.527636584471073\\
-8.399917	0.605177554623764\\
-5.599898	0.628589480563254\\
-2.799895	0.694526376795682\\
0.0001231027	0.808082455888235\\
2.800126	0.699678842282476\\
5.600144	0.641985933927142\\
8.400147	0.633421786584242\\
11.20017	0.561692990300011\\
14.00017	0.39306892050227\\
16.80019	0.317618961670328\\
19.60019	0.250786498912727\\
22.40021	0.224202225268904\\
25.20021	0.210821225263861\\
28.00023	0.203980230352655\\
30.80023	0.383392719912712\\
33.60025	2.39730050410771\\
36.40025	3.46838334214678\\
39.20027	3.43167681157607\\
42.00027	3.78583929167053\\
44.80029	3.75656447311816\\
47.6003	5.61934351808011\\
50.40031	8.62840646100036\\
53.20032	8.97316837826044\\
56.00034	9.16751622747819\\
58.80032	7.01064846154231\\
61.60028	8.44821325947328\\
64.40024	9.26072144129561\\
67.2002	8.71253337962562\\
70.00015	7.98376084836885\\
};
\addlegendentry{Estimator RMSE}

\addplot [color=mycolor2, line width=2.0pt]
  table[row sep=crcr]{%
-70	inf\\
-67.2	inf\\
-64.4	inf\\
-61.60001	inf\\
-58.8	inf\\
-56	inf\\
-53.20001	inf\\
-50.40001	inf\\
-47.60001	inf\\
-44.80002	inf\\
-42.00002	inf\\
-39.20003	132.671060495151\\
-36.40002	34.2167483986679\\
-33.60001	24.8000029265009\\
-30.8	13.5253071429235\\
-27.99998	11.3978103159041\\
-25.19998	0.491381366395962\\
-22.39996	0.389690035049608\\
-19.59996	0.324533907087311\\
-16.79994	0.261242157861812\\
-13.99994	0.223614656275318\\
-11.19992	0.160423235837018\\
-8.399917	0.122575358394569\\
-5.599898	0.0944020906174028\\
-2.799895	0.0756458181188343\\
0.0001231027	0.089681077450975\\
2.800126	0.0832168264601058\\
5.600144	0.0902973986733282\\
8.400147	0.14874742703526\\
11.20017	0.201891146144445\\
14.00017	0.243770944064229\\
16.80019	0.234372003520755\\
19.60019	0.340916675478912\\
22.40021	0.459578946366677\\
25.20021	0.59176506574483\\
28.00023	17.0834194152956\\
30.80023	22.6036848838681\\
33.60025	19.9479048013762\\
36.40025	38.678296357808\\
39.20027	inf\\
42.00027	inf\\
44.80029	inf\\
47.6003	inf\\
50.40031	inf\\
53.20032	inf\\
56.00034	inf\\
58.80032	inf\\
61.60028	inf\\
64.40024	inf\\
67.2002	inf\\
70.00015	inf\\
};
\addlegendentry{All-paths REB}

\addplot [color=mycolor3,line width=2.0pt,dashed]
  table[row sep=crcr]{%
-70	0.0161334438017548\\
-67.2	0.0155051895783237\\
-64.4	0.014879114321494\\
-61.60001	0.0142520415766495\\
-58.8	0.0136262728405533\\
-56	0.0130025065320495\\
-53.20001	0.0123787582250001\\
-50.40001	0.0117564738076016\\
-47.60001	0.0111372272318532\\
-44.80002	0.0105190635924539\\
-42.00002	0.00990323490440181\\
-39.20003	0.0092902431469222\\
-36.40002	0.00868014701042066\\
-33.60001	0.00807564241957423\\
-30.8	0.00747527339117323\\
-27.99998	0.00688219669356511\\
-25.19998	0.0062976291206029\\
-22.39996	0.00572436187531005\\
-19.59996	0.00516567502015202\\
-16.79994	0.00462729386030303\\
-13.99994	0.00411791218146301\\
-11.19992	0.00364818690913255\\
-8.399917	0.00323651094947513\\
-5.599898	0.00290788401854884\\
-2.799895	0.00269263248271725\\
0.0001231027	0.00261683946372077\\
2.800126	0.00269263248271725\\
5.600144	0.00290821882034087\\
8.400147	0.0032368835880188\\
11.20017	0.00364818690913255\\
14.00017	0.00411791218146301\\
16.80019	0.0046278266278642\\
19.60019	0.00516567502015202\\
22.40021	0.00572436187531005\\
25.20021	0.00629762912060289\\
28.00023	0.00688219669356511\\
30.80023	0.00747527339117323\\
33.60025	0.00807564241957423\\
36.40025	0.00868114640680503\\
39.20027	0.0092902431469222\\
42.00027	0.00990323490440181\\
44.80029	0.0105190635924539\\
47.6003	0.0111372272318532\\
50.40031	0.0117564738076016\\
53.20032	0.0123787582250001\\
56.00034	0.0130025065320495\\
58.80032	0.0136262728405533\\
61.60028	0.0142520415766495\\
64.40024	0.014879114321494\\
67.2002	0.0155051895783237\\
70.00015	0.0161334438017548\\
};
\addlegendentry{LoS-only REB}

\addplot [color=mycolor4,line width=2.0pt]
  table[row sep=crcr]{%
% -70	6.39592667321668\\
% -67.2	6.5343259795443\\
% -64.4	6.33462083934746\\
% -61.60001	3.57262830513547\\
% -58.8	3.6948977369303\\
% -56	3.82505113188721\\
% -53.20001	3.91058082406948\\
% -50.40001	3.94401273646638\\
% -47.60001	3.96557353652529\\
% -44.80002	3.9774428687584\\
% -42.00002	4.0378426739229\\
% -39.20003	4.12654936826338\\
% -36.40002	4.15320786062927\\
% -33.60001	3.19353189479216\\
% -30.8	0.591666162165734\\
% -27.99998	0.209152024325448\\
% -25.19998	0.172738402342477\\
% -22.39996	0.205628112841815\\
% -19.59996	0.243929252768003\\
% -16.79994	0.288438766476303\\
% -13.99994	0.33970281777764\\
% -11.19992	0.397080825313864\\
% -8.399917	0.457786538497459\\
% -5.599898	0.515120864725034\\
% -2.799895	0.557898150005047\\
% 0.0001231027	0.573958001687763\\
% 2.800126	0.557895309325265\\
% 5.600144	0.51511746733714\\
% 8.400147	0.457819445994653\\
% 11.20017	0.397045936789452\\
% 14.00017	0.339698090089848\\
% 16.80019	0.288463842643309\\
% 19.60019	0.243909900578505\\
% 22.40021	0.205630399805376\\
% 25.20021	0.172740597656387\\
% 28.00023	0.415248246628177\\
% 30.80023	0.754177142280435\\
% 33.60025	3.25734727127905\\
% 36.40025	4.08020601590652\\
% 39.20027	4.09634023646321\\
% 42.00027	4.02092377436027\\
% 44.80029	3.96076285831971\\
% 47.6003	3.95199589670662\\
% 50.40031	3.9331607897931\\
% 53.20032	3.90288048062415\\
% 56.00034	3.81687410949333\\
% 58.80032	3.68895466811549\\
% 61.60028	3.56722235163668\\
% 64.40024	6.33546802063105\\
% 67.2002	6.53250474166671\\
% 70.00015	6.40861044703423\\
% };
-70	6.40206342462915\\
-67.2	6.58734938497596\\
-64.4	6.77447096303103\\
-61.60001	6.96672717421591\\
-58.8	7.16226638886223\\
-56	7.36983841055953\\
-53.20001	7.21526613637536\\
-50.40001	7.46403601707845\\
-47.60001	6.35234558968988\\
-44.80002	4.76856660355872\\
-42.00002	4.41848809693007\\
-39.20003	4.36187639838291\\
-36.40002	4.31410026073952\\
-33.60001	3.19353085569978\\
-30.8	0.591667720285662\\
-27.99998	0.209144692253023\\
-25.19998	0.308088776445678\\
-22.39996	0.345542007295323\\
-19.59996	0.364880560454831\\
-16.79994	0.373013516813337\\
-13.99994	0.339703399965663\\
-11.19992	0.397080446986194\\
-8.399917	0.457785655743199\\
-5.599898	0.515122649904995\\
-2.799895	0.55789760677327\\
0.0001231027	0.573957895747445\\
2.800126	0.557896122123204\\
5.600144	0.515116690068291\\
8.400147	0.457814979041739\\
11.20017	0.397046004424952\\
14.00017	0.339697548397442\\
16.80019	0.371699241295597\\
19.60019	0.360193025576599\\
22.40021	0.345574092461229\\
25.20021	0.348840515331774\\
28.00023	0.415247843541976\\
30.80023	0.754174256033195\\
33.60025	3.2573471956597\\
36.40025	4.12555276351184\\
39.20027	4.17884527059336\\
42.00027	4.27318956476036\\
44.80029	4.66949062271838\\
47.6003	6.31494862869126\\
50.40031	7.53924757663\\
53.20032	7.23441716005735\\
56.00034	7.3998208543894\\
58.80032	7.18774632291336\\
61.60028	6.98279540641802\\
64.40024	6.79131401861698\\
67.2002	6.59896088197674\\
70.00015	6.41472300097381\\
};
\addlegendentry{WAA REB}

\end{axis}

\end{tikzpicture}%
\caption{Scenario 1 (bicycle): Comparison among the \ac{RMSE} of the estimated range between the bicycle and the \ac{RSU} and three \acp{REB}.}
\label{Fig.bike}
\vspace{-4mm} 
\end{figure}
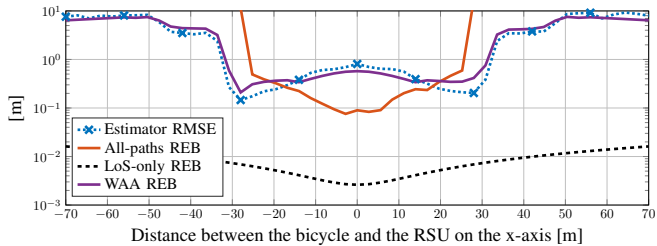

\subsection{Results and Discussion}
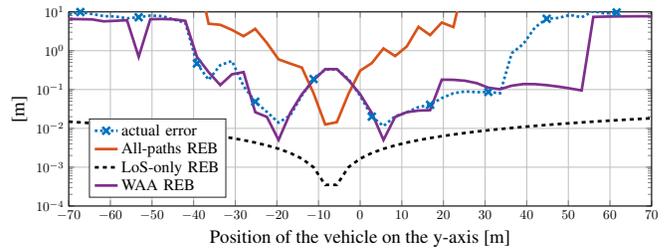
\begin{figure}
\center
\definecolor{mycolor1}{rgb}{0.00000,0.44700,0.74100}%
\definecolor{mycolor2}{rgb}{0.85000,0.32500,0.09800}%
\definecolor{mycolor3}{rgb}{0,0,0}%
\definecolor{mycolor4}{rgb}{0.49400,0.18400,0.55600}%
\begin{tikzpicture}[scale=0.8\linewidth/14cm]

\begin{axis}[%
width=6.028in,
height=2.009in,
scale only axis,
unbounded coords=jump,
xmin=-70,
xmax=70,
xlabel style={font=\color{white!15!black},font=\Large},
xlabel={Position of the vehicle on the y-axis [m]},
ymode=log,
ymin=0.0001,
ymax=10,
ylabel style={font=\color{white!15!black},font=\Large},
ylabel={[m]},
yminorticks=true,
%axis x line*=bottom,
%axis y line*=left,
xmajorgrids,
ymajorgrids,
axis background/.style={fill=white},
legend style={at={(0.27,0.45)},legend cell align=left, align=left, draw=white!15!black,font=\large}
]
\addplot [color=mycolor1, dotted, line width=2.0pt,mark=x, mark repeat=5,mark options={solid},mark size=4pt]
  table[row sep=crcr]{%
-70.000002162058	8.73374042029636\\
-67.199999	9.90553056971648\\
-64.400004	8.85402641689555\\
-61.6	7.80194497682694\\
-58.8	7.61991560114187\\
-56	6.83177837063112\\
-53.2	7.28313582295184\\
-50.4	8.04699269617751\\
-47.6	7.82090074254828\\
-44.8	6.38186474284122\\
-42	5.22869663030207\\
-39.2	0.469572159411299\\
-36.40001	0.177147862718021\\
-33.60001	0.424748940328811\\
-30.80001	0.544934407620232\\
-28	0.122662367484436\\
-25.2	0.0485543185332299\\
-22.4	0.0265526632198271\\
-19.59999	0.0140474929260164\\
-16.79999	0.021553069847098\\
-13.99999	0.0704179741754174\\
-11.19999	0.18677583329996\\
-8.39998	0.324875991388594\\
-5.59998000000002	0.336664403843236\\
-2.79997	0.186177105861708\\
2.99999999811007e-05	0.0696388497956418\\
2.80002999999999	0.0200462037939835\\
5.60004000000001	0.0115099226941389\\
8.40003999999999	0.0204502738264581\\
11.20004	0.0260213961389768\\
14.00005	0.0352294432838416\\
16.80005	0.0405446869414517\\
19.60005	0.060256636568261\\
22.40005	0.080271019218589\\
25.20006	0.0892844870002146\\
28.00006	0.085540730808858\\
30.8001	0.0868194760454677\\
33.6001	0.0802249228861941\\
36.4001	0.856979920799201\\
39.2001	1.4988133780224\\
42.0001	4.00589252443792\\
44.8001	6.66485472354433\\
47.6001	6.9890977915914\\
50.4001	8.36497969465309\\
53.2001	7.26877083809742\\
56.0001	10.463500555309\\
58.8001	9.08479344363106\\
61.6001	9.65841954702263\\
64.4001	14.3119379440696\\
67.2001	11.3173648580808\\
70	11.2149005002578\\
};
\addlegendentry{actual error}

\addplot [color=mycolor2, line width=2.0pt]
  table[row sep=crcr]{%
-70.000002162058	inf\\
-67.199999	inf\\
-64.400004	inf\\
-61.6	inf\\
-58.8	1425.33073659082\\
-56	inf\\
-53.2	inf\\
-50.4	inf\\
-47.6	355.463184409492\\
-44.8	inf\\
-42	inf\\
-39.2	118.743434999762\\
-36.40001	4.96100134795955\\
-33.60001	4.65286738645768\\
-30.80001	3.44148414766009\\
-28	2.36681808522808\\
-25.2	3.6594984045669\\
-22.4	1.58538105991954\\
-19.59999	0.599157299786867\\
-16.79999	0.466820389428998\\
-13.99999	0.368029265091485\\
-11.19999	0.0809971219428059\\
-8.39998	0.0125870437239775\\
-5.59998000000002	0.0144186760119486\\
-2.79997	0.0779755918661739\\
2.99999999811007e-05	0.306595082872571\\
2.80002999999999	0.479649588169444\\
5.60004000000001	1.14544890907344\\
8.40003999999999	0.728160309247039\\
11.20004	1.28909992977695\\
14.00005	4.1193337352047\\
16.80005	2.52351777285462\\
19.60005	5.31728611896634\\
22.40005	4.07086743417123\\
25.20006	76.3143409370106\\
28.00006	inf\\
30.8001	inf\\
33.6001	inf\\
36.4001	inf\\
39.2001	inf\\
42.0001	inf\\
44.8001	inf\\
47.6001	inf\\
50.4001	inf\\
53.2001	inf\\
56.0001	inf\\
58.8001	inf\\
61.6001	inf\\
64.4001	inf\\
67.2001	inf\\
70	inf\\
};
\addlegendentry{All-paths REB}

\addplot [color=mycolor3, line width=2.0pt, dashed]
  table[row sep=crcr]{%
-70.000002162058	0.014903115958115\\
-67.199999	0.0142405604151647\\
-64.400004	0.0135792905437373\\
-61.6	0.0129159715766682\\
-58.8	0.012253977557869\\
-56	0.0115911648018553\\
-53.2	0.0106862930737136\\
-50.4	0.0102666103765469\\
-47.6	0.00960455130166616\\
-44.8	0.00894287335731615\\
-42	0.00827993792018502\\
-39.2	0.00761775639841677\\
-36.40001	0.00695547959085219\\
-33.60001	0.00629328037469913\\
-30.80001	0.00563089048212728\\
-28	0.00496909249290882\\
-25.2	0.00430702190583903\\
-22.4	0.00364482835172546\\
-19.59999	0.00298282023504321\\
-16.79999	0.00232101036885695\\
-13.99999	0.00165988148485102\\
-11.19999	0.00100017750044432\\
-8.39998	0.000350531523672047\\
-5.59998000000002	0.000350531523672047\\
-2.79997	0.00100017750044432\\
2.99999999811007e-05	0.00165988148485102\\
2.80002999999999	0.00232101036885695\\
5.60004000000001	0.00298282023504321\\
8.40003999999999	0.00364482835172546\\
11.20004	0.00430702190583903\\
14.00005	0.00496909249290882\\
16.80005	0.00563089048212728\\
19.60005	0.00629328037469914\\
22.40005	0.00695547959085219\\
25.20006	0.00761775639841677\\
28.00006	0.00827993792018501\\
30.8001	0.00894287335731615\\
33.6001	0.00960455130166616\\
36.4001	0.0102666103765469\\
39.2001	0.0109289158150324\\
42.0001	0.0115911648018553\\
44.8001	0.012253977557869\\
47.6001	0.0129159715766682\\
50.4001	0.0135792905437373\\
53.2001	0.0142405604151647\\
56.0001	0.014903115958115\\
58.8001	0.0155660019537509\\
61.6001	0.0162284514512292\\
64.4001	0.0168899000921978\\
67.2001	0.0175520190108966\\
70	0.0182149121344672\\
};
\addlegendentry{LoS-only REB}

\addplot [color=mycolor4, line width=2.0pt]
  table[row sep=crcr]{%
% -70.000002162058	0.20729523944188\\
% -67.199999	0.301747843112816\\
% -64.400004	0.45311280675992\\
% -61.6	2.19165440646551\\
% -58.8	2.63282690314906\\
% -56	3.08869395144715\\
% -53.2	0.727161367600034\\
% -50.4	4.10170236992999\\
% -47.6	4.43920241959111\\
% -44.8	4.63743136879873\\
% -42	4.58777333632249\\
% -39.2	0.69340949982651\\
% -36.40001	0.0295745092289916\\
% -33.60001	0.0300532557874219\\
% -30.80001	0.0299901264417202\\
% -28	0.0291709134477883\\
% -25.2	0.0264856906894078\\
% -22.4	0.0200480939058063\\
% -19.59999	0.00584806514609808\\
% -16.79999	0.0249803757628013\\
% -13.99999	0.075763146658572\\
% -11.19999	0.174175972569871\\
% -8.39998	0.335359077022036\\
% -5.59998000000002	0.335355155029393\\
% -2.79997	0.174174803870312\\
% 2.99999999811007e-05	0.0757623776302444\\
% 2.80002999999999	0.0249841549086869\\
% 5.60004000000001	0.00584159089331155\\
% 8.40003999999999	0.0200457890866149\\
% 11.20004	0.0264929700744977\\
% 14.00005	0.0291654690818564\\
% 16.80005	0.0299988942855651\\
% 19.60005	0.0300861268043879\\
% 22.40005	0.0296802079850087\\
% 25.20006	0.168736363711666\\
% 28.00006	0.146310198390899\\
% 30.8001	0.111934817546339\\
% 33.6001	0.102261873473445\\
% 36.4001	0.126158471724396\\
% 39.2001	0.139161134936694\\
% 42.0001	0.138490092555513\\
% 44.8001	0.130993693637672\\
% 47.6001	0.119993345069264\\
% 50.4001	0.108185366132816\\
% 53.2001	0.0965035347777683\\
% 56.0001	0.0858229924332498\\
% 58.8001	0.0753516954408118\\
% 61.6001	0.0658493592610292\\
% 64.4001	0.0583690537503261\\
% 67.2001	0.0516812663526271\\
% 70	0.0465325054505562\\
% };
-70.000002162058	6.56495516866777\\
-67.199999	6.49742081730843\\
-64.400004	6.41364630370138\\
-61.6	5.71579054924543\\
-58.8	5.89744121858398\\
-56	6.10895565473955\\
-53.2	0.696267659555085\\
-50.4	6.49114098387214\\
-47.6	6.56604032920814\\
-44.8	6.43022838886761\\
-42	5.88720942997685\\
-39.2	0.69335809491905\\
-36.40001	0.280716980325252\\
-33.60001	0.130363672415016\\
-30.80001	0.247374610444757\\
-28	0.281330055469485\\
-25.2	0.0261225009991602\\
-22.4	0.0197113118014959\\
-19.59999	0.00502811479822815\\
-16.79999	0.024872969204514\\
-13.99999	0.075744275568562\\
-11.19999	0.174172541661542\\
-8.39998	0.335358883102472\\
-5.59998000000002	0.335354954062469\\
-2.79997	0.174171310365254\\
2.99999999811007e-05	0.0757438787829798\\
2.80002999999999	0.0248728670083563\\
5.60004000000001	0.00502810424644265\\
8.40003999999999	0.0197112886325383\\
11.20004	0.0261224940035686\\
14.00005	0.0286984543627383\\
16.80005	0.0294197927391448\\
19.60005	0.179599566513464\\
22.40005	0.178070641653955\\
25.20006	0.168520499334303\\
28.00006	0.146020570832991\\
30.8001	0.11139666627412\\
33.6001	0.101707222522245\\
36.4001	0.125609557074472\\
39.2001	0.138461495002674\\
42.0001	0.137677023939431\\
44.8001	0.129868998005677\\
47.6001	0.118977408769234\\
50.4001	0.106936492901082\\
53.2001	0.0947703193627151\\
56.0001	7.43603013423632\\
58.8001	7.56026338697823\\
61.6001	7.65603323295901\\
64.4001	7.70298979429468\\
67.2001	7.73318798970938\\
70	7.723088900799\\
};
\addlegendentry{WAA REB}

\end{axis}
\end{tikzpicture}%
\caption{Scenario 2: Comparison among the \ac{RMSE} of the estimated range between the vehicle and the bicycle and three \acp{REB}.}
\label{Fig.mov}
\vspace{-4mm} 
\end{figure}

%\subsubsection{Scenario 1}
%We firstly analyze how \acp{RMSE} and three \acp{REB} change with positions of the vehicle and bicycle in the first scenario, as illustrated in Fig.~\ref{car} and Fig.~\ref{Fig.bike}, respectively.  
Fig.~\ref{car} and Fig.~\ref{Fig.bike} show the RMSE performance of the ranging to the vehicle and bicycle, respectively for Scenario 1.  Fig.~\ref{Fig.mov} shows the \ac{RMSE} for Scenario 2 (without RSU). With respect to the positioning requirements, we note that for Scenario 1, within about 40 m from the RSU, the moderate accuracy (1--3 m) requirement can be exceeded. For Scenario 2, moderate accuracy is also attainable when vehicle and bicycle are sufficiently close to each-other.  

Since the three figures exhibit a similar trend, we focus our  discussion on Fig.~\ref{car}. First of all, we note that 
%The results demonstrate that we can acquire below $1\,\text{m}$ range error at the intersection, as blue lines are lower than $1\,\text{m}$ when $y \in [-20\,\text{m} , 20\,\text{m}]$ in Fig.~\ref{car} and $x \in [-20\,\text{m} , 20\,\text{m}]$ in Fig.~\ref{Fig.bike}. Both figures also show that 
the WAA \ac{REB} is more reasonable than the \ac{LoS} REB and the all-paths \ac{REB}. This is because the \ac{LoS}-only \ac{REB} only considers the \ac{LoS} path, which is overly optimistic, while the all-paths \ac{REB} assumes all paths are resolvable, which  leads to
unreasonable results. 
The two bounds differ with 1--3 orders of magnitude, where the LoS-only REB promises sub-cm accuracy, while the all-paths REB varies from infinite error down to about 10 cm. On the other hand,  the proposed WAA \acp{REB} takes the path non-resolvability into consideration and matches the estimator's performance well. 
Secondly, we observe that the  \ac{LoS} and all-paths \acp{REB} decrease with the \ac{UE} approaching the \ac{RSU}, because the \ac{LoS} path increases in power.  %The two bounds differ with 1-3 orders of magnitude, where the LoS-only bound promises sub-cm accuracy, while the all-path bounds varies from infinite error down to about 10 cm. The estimator RMSE and the WAA REB have comparable trends .-
In contrast, the algorithm and the WAA REB exhibit different behavior: around  $y=-39.2\, \text{m}$ in Fig.~\ref{car}, the  WAA \acp{REB} rapidly drops when approaching the RSU. This effect is caused by the sudden 
%The sudden decreases/increases are caused by the 
disappearance of strong \ac{NLoS} paths caused by the buildings, which in turn  alleviates the inter-path interference. For example, the vehicle always has strong reflection paths from the buildings C and D at very beginning, but the buildings C and D cannot reflect any paths to vehicles anymore after $[1.6 \, \text{m},-39.2 \, \text{m},1.5\, \text{m}]^{\mathsf{T}}$, and the rest \ac{NLoS} paths are relatively weak, thus the \ac{LoS} becomes more dominate and the WAA \ac{REB} drops accordingly. Later, when the vehicle starts to receive reflections from buildings A and B, the WAA \ac{REB}  will have a sudden increase. A second interesting effect appears when the the vehicle is in close proximity to the RSU (for $|y|<30\, \text{m}$). As the vehicle approaches the RSU, the WAA increases, in sharp contrast to the LoS-only and all-paths REBs. This effect is due to the ground reflection. Because of the limited bandwidth, the ground reflection always falls in the LoS resolution cell. Close to the RSU, the ground reflection is relatively strong with an amplitude of around 50\% of the LoS path and an excess delay over $1~\text{m}$. Combined, this leads to a significant bias. This result shows that multiple antennas should be deployed to suppress ground reflections when bandwidth is limited.

%\subsubsection{Scenario 2}
%We then analyze the performance of the range estimation for the second scenario, and compare \acp{RMSE} with three \acp{REB}, as shown in Fig.~\ref{Fig.mov}. From the figure, we find that the \ac{RMSE} is below  $0.5\,\text{m}$ in this scenario, when the vehicle enter the intersection. Similar as the first scenario, the merged \ac{REB} fits the \ac{RMSE} better, compared to the \ac{LoS} and the \ac{NLoS} \acp{REB}, as the magenta is the closet to the blue line. This is due to the merged \ac{REB} considers multipath and the path non-resolvability. We also observe that before the collision, overall, the merged \ac{REB} firstly goes up because of the growing \ac{NLoS} paths, then drops due to the loss of some strong \ac{NLoS} paths from the building D, and then grows again since the \ac{NLoS} paths grow faster.

\section{Conclusions} \label{conclusion}
In this paper, we focus on the analysis of sidelink V2X positioning towards 3GPP Release 18 in sub-6 GHz, where a novel methodology based on Fisher information analysis to predict positioning performance in a multipath propagation environment is provided, and the sidelink positioning and the method for positioning performance prediction are evaluated for two common urban scenarios occurring at an intersection using ray-tracing data. Our results indicate that the proposed WAA \ac{REB} can  predict the positioning performance better than the conventional bounds from the literature, since both multipath and the path non-resolvability are considered. The results indicate that biases due to multipath are the main cause of the error. Nevertheless, sub-meter accuracy was achievable, even in a complex urban scenario, when the transmitter and the receiver are sufficiently close. The impact of ground reflection was studied and found to be an important error source. 

%The results also confirm that the merged \ac{REB} is lower so that we can have better positioning performance,  when  multipath becomes simpler and inter-path interference becomes milder.

\section{Acknowledgments}
This work was partially supported by MSCA-IF grant 888913 (OTFS-RADCOM), and by the Wallenberg AI, Autonomous Systems and Software Program (WASP) funded by Knut and Alice Wallenberg Foundation. The authors wish to thank Remcom for providing Wireless InSite\textregistered  ray-tracer.

%\section{Literature}

%\begin{itemize}
%\item 3GPP; Technical Specification Group Radio Access Network; Study on scenarios and requirements of in-coverage, partial coverage, and out-of-coverage NR positioning use cases (Release 17)

%\item Ranging and velocity estimation with sidelink 5G V2X is investigated in \cite{JCS_V2X_2022} using both CRB analysis and simulation results.
%\end{itemize}
\balance
\bibliography{IEEEabrv_paper,Bibl_paper}

\begin{thebibliography}{10}
\providecommand{\url}[1]{#1}
\csname url@rmstyle\endcsname
\providecommand{\newblock}{\relax}
\providecommand{\bibinfo}[2]{#2}
\providecommand\BIBentrySTDinterwordspacing{\spaceskip=0pt\relax}
\providecommand\BIBentryALTinterwordstretchfactor{4}
\providecommand\BIBentryALTinterwordspacing{\spaceskip=\fontdimen2\font plus
\BIBentryALTinterwordstretchfactor\fontdimen3\font minus
  \fontdimen4\font\relax}
\providecommand\BIBforeignlanguage[2]{{%
\expandafter\ifx\csname l@#1\endcsname\relax
\typeout{** WARNING: IEEEtran.bst: No hyphenation pattern has been}%
\typeout{** loaded for the language `#1'. Using the pattern for}%
\typeout{** the default language instead.}%
\else
\language=\csname l@#1\endcsname
\fi
#2}}
\renewcommand\BIBentryALTinterwordstretchfactor{4}

\bibitem{wild2021joint}
T.~Wild, V.~Braun, and H.~Viswanathan, ``Joint design of communication and
  sensing for beyond {5G} and {6G} systems,'' \emph{IEEE Access}, vol.~9, pp.
  30\,845--30\,857, 2021.

\bibitem{KanRap:21}
O.~{Kanhere} and T.~S. {Rappaport}, ``Position location for futuristic cellular
  communications: {5G} and beyond,'' \emph{{IEEE} Communications Magazine},
  vol.~59, no.~1, pp. 70--75, 2021.

\bibitem{hexax_d31}
\BIBentryALTinterwordspacing
H.~Wymeersch \emph{et~al.}, ``Localisation and sensing use cases and gap
  analysis,'' Hexa-X project Deliverable D3.1, v1.4, 2022. [Online]. Available:
  \url{https://hexa-x.eu/deliverables/}
\BIBentrySTDinterwordspacing

\bibitem{tr:38859-3gpp22}
3GPP, ``Study on expanded and improved {NR} positioning,'' TR 38.859, Technical
  Report 0.1.0, 2022.

\bibitem{tr:38845-3gpp21}
3GPP, ``Study on scenarios and requirements of in-coverage, partial coverage,
  and out-of-coverage {NR} positioning use cases,'' TR 38.845, Technical Report
  17.0.0, 2021.

\bibitem{ko2021v2x}
S.-W. Ko, H.~Chae, K.~Han, S.~Lee, D.-W. Seo, and K.~Huang, ``{V2X}-based
  vehicular positioning: Opportunities, challenges, and future directions,''
  \emph{IEEE Wireless Communications}, vol.~28, no.~2, pp. 144--151, 2021.

\bibitem{bartoletti2021positioning}
S.~Bartoletti, H.~Wymeersch, T.~Mach, O.~Brunneg{\aa}rd, D.~Giustiniano,
  P.~Hammarberg, M.~F. Keskin, J.~O. Lacruz, S.~M. Razavi, J.~R{\"o}nnblom,
  \emph{et~al.}, ``Positioning and sensing for vehicular safety applications in
  {5G} and beyond,'' \emph{IEEE Communications Magazine}, vol.~59, no.~11, pp.
  15--21, 2021.

\bibitem{saily2021positioning}
M.~S{\"a}ily, O.~N. Yilmaz, D.~S. Michalopoulos, E.~P{\'e}rez, R.~Keating, and
  J.~Schaepperle, ``Positioning technology trends and solutions toward {6G},''
  in \emph{IEEE International Symposium on Personal, Indoor and Mobile Radio
  Communications (PIMRC)}, 2021.

\bibitem{lu2021joint}
Y.~Lu, M.~Koivisto, J.~Talvitie, E.~Rastorgueva-Foi, T.~Levanen, E.~S. Lohan,
  and M.~Valkama, ``Joint positioning and tracking via {NR} sidelink in
  {5G}-empowered industrial {IoT}: Releasing the potential of {V2X}
  technology,'' \emph{arXiv preprint arXiv:2101.06003}, 2021.

\bibitem{bazzi2019survey}
A.~Bazzi, G.~Cecchini, M.~Menarini, B.~M. Masini, and A.~Zanella, ``Survey and
  perspectives of vehicular {Wi-Fi} versus sidelink cellular-{V2X} in the {5G}
  era,'' \emph{Future Internet}, vol.~11, no.~6, p. 122, 2019.

\bibitem{liu2021highly}
Q.~Liu, P.~Liang, J.~Xia, T.~Wang, M.~Song, X.~Xu, J.~Zhang, Y.~Fan, and
  L.~Liu, ``A highly accurate positioning solution for {C-V2X} systems,''
  \emph{Sensors}, vol.~21, no.~4, p. 1175, 2021.

\bibitem{del2017performance}
J.~A. del Peral-Rosado, M.~A. Barreto-Arboleda, F.~Zanier, G.~Seco-Granados,
  and J.~A. L{\'o}pez-Salcedo, ``Performance limits of {V2I} ranging
  localization with {LTE} networks,'' in \emph{IEEE Workshop on Positioning,
  Navigation and Communications (WPNC)}, 2017.

\bibitem{hossain2017high}
M.~A. Hossain, I.~Elshafiey, and A.~Al-Sanie, ``High accuracy {GPS}-free
  vehicular positioning based on {V2V} communications and {RSU}-assisted {DOA}
  estimation,'' in \emph{IEEE-GCC Conference and Exhibition}, 2017.

\bibitem{kakkavas2018multi}
A.~Kakkavas, M.~H.~C. Garcia, R.~A. Stirling-Gallacher, and J.~A. Nossek,
  ``Multi-array {5G V2V} relative positioning: Performance bounds,'' in
  \emph{IEEE Global Communications Conference (GLOBECOM)}, 2018, pp. 206--212.

\bibitem{JCS_V2X_2022}
N.~Decarli, S.~Bartoletti, and B.~M. Masini, ``Joint communication and sensing
  in {5G-V2X} vehicular networks,'' in \emph{IEEE Mediterranean
  Electrotechnical Conference (MELECON)}, 2022, pp. 295--300.

\bibitem{ma2019efficient}
S.~Ma, F.~Wen, X.~Zhao, Z.-M. Wang, and D.~Yang, ``An efficient {V2X} based
  vehicle localization using single {RSU} and single receiver,'' \emph{IEEE
  Access}, vol.~7, pp. 46\,114--46\,121, 2019.

\bibitem{liu2021v2x}
Q.~Liu, R.~Liu, Z.~Wang, L.~Han, and J.~S. Thompson, ``A {V2X}-integrated
  positioning methodology in ultradense networks,'' \emph{IEEE Internet of
  Things Journal}, vol.~8, no.~23, pp. 17\,014--17\,028, 2021.

\bibitem{fouda2021dynamic}
A.~Fouda, R.~Keating, and A.~Ghosh, ``Dynamic selective positioning for
  high-precision accuracy in {5G NR} {V2X} networks,'' in \emph{IEEE Vehicular
  Technology Conference (VTC2021-Spring)}, 2021.

\bibitem{5GAA-SysArch}
\BIBentryALTinterwordspacing
``System architecture and solution development; high-accuracy positioning for
  {C-V2X},'' 5GAA Automotive Association Technical Report, 2021. [Online].
  Available: \url{http://www.5gaa.org}
\BIBentrySTDinterwordspacing

\bibitem{tr:37885-3gpp19}
3GPP, ``Technical specification group radio access network; study on evaluation
  methodology of new vehicle-to-everything ({V2X}) use cases for {LTE} and
  {NR},'' TR 37.885, Technical Report 15.3.0, 2019.

\bibitem{Passive_OFDM_2010}
C.~R. Berger, B.~Demissie, J.~Heckenbach, P.~Willett, and S.~Zhou, ``Signal
  processing for passive radar using {OFDM} waveforms,'' \emph{IEEE Journal of
  Selected Topics in Signal Processing}, vol.~4, no.~1, pp. 226--238, 2010.

\bibitem{behravan-6g-2022}
A.~Behravan, V.~Yajnanarayana, M.~F. Keskin, H.~Chen, T.~E.~A. Deep~Shrestha,
  T.~Svensson, K.~Schindhelm, A.~Wolfgang, S.~Lindberg, and H.~Wymeersch,
  ``Positioning and sensing in {6G}: Gaps, challenges, and opportunities,''
  \emph{IEEE Vehicular Technology Magazine (under review)}, 2022.

\bibitem{VanTrees}
H.~L.~V. Trees, \emph{Detection, Estimation, and Modulation Theory}.\hskip 1em
  plus 0.5em minus 0.4em\relax John Wiley \& Sons, New York, 2004.

\bibitem{shen_localization}
Y.~Shen and M.~Z. Win, ``Fundamental limits of wideband localization— part
  {I}: A general framework,'' \emph{IEEE Transactions on Information Theory},
  vol.~56, no.~10, pp. 4956--4980, 2010.

\bibitem{OFDM_Radar_Corr_TAES_2020}
S.~Mercier, S.~Bidon, D.~Roque, and C.~Enderli, ``Comparison of
  correlation-based {OFDM} radar receivers,'' \emph{IEEE Transactions on
  Aerospace and Electronic Systems}, vol.~56, no.~6, pp. 4796--4813, 2020.

\bibitem{RadCom_Proc_IEEE_2011}
C.~{Sturm} and W.~{Wiesbeck}, ``Waveform design and signal processing aspects
  for fusion of wireless communications and radar sensing,'' \emph{Proceedings
  of the IEEE}, vol.~99, no.~7, pp. 1236--1259, July 2011.

\bibitem{OFDM_Radar_Phd_2014}
M.~Braun, ``{OFDM} radar algorithms in mobile communication networks,''
  \emph{Karlsruher Institutes f{\"u}r Technologie}, 2014.

\bibitem{toa_multipath}
Y.~Qi, H.~Kobayashi, and H.~Suda, ``On time-of-arrival positioning in a
  multipath environment,'' \emph{IEEE Transactions on Vehicular Technology},
  vol.~55, no.~5, pp. 1516--1526, 2006.

\bibitem{5g_nr_rel16}
S.~Parkvall, Y.~Blankenship, R.~Blasco, E.~Dahlman, G.~Fodor, S.~Grant,
  E.~Stare, and M.~Stattin, ``{5G NR} release 16: Start of the {5G}
  evolution,'' \emph{IEEE Communications Standards Magazine}, vol.~4, no.~4,
  pp. 56--63, 2020.

\bibitem{5g_nr_v2x_driving}
H.~Bagheri, M.~Noor-A-Rahim, Z.~Liu, H.~Lee, D.~Pesch, K.~Moessner, and
  P.~Xiao, ``{5G NR-V2X}: Toward connected and cooperative autonomous
  driving,'' \emph{IEEE Communications Standards Magazine}, vol.~5, no.~1, pp.
  48--54, 2021.

\bibitem{zhu2009simple}
S.~Zhu and Z.~Ding, ``A simple approach of range-based positioning with low
  computational complexity,'' \emph{IEEE Transactions on Wireless
  Communications}, vol.~8, no.~12, pp. 5832--5836, 2009.

\bibitem{WirelessInSite}
\BIBentryALTinterwordspacing
{Remcom}. {Wireless {InSite} - {3D} Wireless Prediction Software}. Accessed:
  Oct 9, 2022). [Online]. Available:
  \url{https://www.remcom.com/wireless-insite-em-propagation-software}
\BIBentrySTDinterwordspacing

\end{thebibliography}

\end{document}